\documentclass[11pt,a4paper]{emulateapj}

\usepackage{psfrag}
\usepackage{psfig}
\usepackage{pspicture}
\usepackage{graphicx}

\usepackage{amssymb,amsmath}
\usepackage{natbib}

\def\apj{ApJ}

\submitted{submitted to \apj}

\begin{document}

\title {Cosmic Web of Galaxies in the COSMOS Field: \\
Public Catalog and Different Quenching for Centrals and Satellites}

\author{
Behnam Darvish\altaffilmark{1},
Bahram Mobasher\altaffilmark{2},
D. Christopher Martin\altaffilmark{1},
David Sobral\altaffilmark{3,4},
Nick Scoville\altaffilmark{1},
Andra Stroe\altaffilmark{5},
Shoubaneh Hemmati\altaffilmark{6},
and Jeyhan Kartaltepe\altaffilmark{7}
}

\setcounter{footnote}{0}
\altaffiltext{1}{Cahill Center for Astrophysics, California Institute of Technology, 1216 East California Boulevard, Pasadena, CA 91125, USA; email: bdarv@caltech.edu, bdarv001@ucr.edu}
\altaffiltext{2}{University of California, Riverside, 900 University Ave, Riverside, CA 92521, USA}
\altaffiltext{3}{Department of Physics, Lancaster University, Lancaster, LA1 4YB, UK}
\altaffiltext{4}{Leiden Observatory, Leiden University, P.O. Box 9513, NL-2300 RA Leiden, The Netherlands}
\altaffiltext{5}{European Southern Observatory, Karl-Schwarzschild-Str. 2, 85748, Garching, Germany}
\altaffiltext{6}{Infrared Processing and Analysis Center, California Institute of Technology, Pasadena, CA 91125, USA}
\altaffiltext{7}{School of Physics and Astronomy, Rochester Institute of Technology, 84 Lomb Memorial Drive, Rochester, NY 14623, USA}

\begin{abstract}
We use a mass complete (log($M/M_{\odot}$) $\geqslant$ 9.6) sample of galaxies with accurate photometric redshifts in the COSMOS field to construct the density field and the cosmic web to $z$=1.2. The comic web extraction relies on the density field Hessian matrix and breaks the density field into clusters, filaments and the field. We provide the density field and cosmic web measures to the community. We show that at $z$ $\lesssim$ 0.8, the median star-formation rate (SFR) in the cosmic web gradually declines from the field to clusters and this decline is especially sharp for satellites ($\sim$ 1 dex vs. $\sim$ 0.5 dex for centrals). However, at $z$ $\gtrsim$ 0.8, the trend flattens out for the overall galaxy population and satellites. For star-forming galaxies only, the median SFR is constant at $z$ $\gtrsim$ 0.5 but declines by $\sim$ 0.3-0.4 dex from the field to clusters for satellites and centrals at $z$ $\lesssim$ 0.5. We argue that for satellites, the main role of the cosmic web environment is to control their star-forming fraction, whereas for centrals, it is mainly to control their overall SFR at $z$ $\lesssim$ 0.5 and to set their fraction at $z$ $\gtrsim$ 0.5. We suggest that most satellites experience a rapid quenching mechanism as they fall from the field into clusters through filaments, whereas centrals mostly undergo a slow environmental quenching at $z$ $\lesssim$ 0.5 and a fast mechanism at higher redshifts. Our preliminary results highlight the importance of the large-scale cosmic web on galaxy evolution.
\end{abstract}


\section{Introduction} \label{intro1}

The standard model of cosmology is based on the cosmological principle, the concept of a spatially homogeneous and isotropic universe when averaged over scales of $\gtrsim$ 100 Mpc. On smaller scales, the universe is inhomogeneous. Dark matter, gas, and galaxies are organized in a complex network known as the cosmic web \citep{Bond96}, which is a direct consequence of the anisotropic gravitational collapse of matter from the early seeds of primordial matter fluctuations \citep{Zeldovich70}. The cosmic web has a broad dynamical range of environments over different physical scales and densities: voids that are deprived of matter and occupy much of the volume of the web, planar walls and sheets, filamentary structures that form at the intersection of walls, and dense clusters and groups of galaxies woven together by filaments. This large-scale picture of the universe has been revealed in numerical simulations and observed distribution of galaxies in the local universe \citep{Davis85,Geller89,Bond96,Colless03,Jarrett04,Doroshkevich04,Jones09,Alpaslan14}. Galaxies form and evolve in the cosmic web and their evolution should be essentially driven by a combination of internal and external processes.
 
Filaments make the backbone of the cosmic web, comprising $\sim$ 40\% of the total mass in the local universe \citep{Aragon-Calvo10}, presumably containing a large fraction of missing baryons in the form of a warm-hot intergalactic medium (IGM) gas \citep{Cen99,Dave01,Shull12,Haider16}, and potentially hosting much of the star-formation activity in the universe \citep{Snedden16}. Recent models of galaxy formation heavily rely on the cold gas flow into galaxies through streams of filaments \citep{Keres05,Dekel09a}, with recent observational evidence supporting this picture \citep{Cantalupo14,Martin14a,Martin14b,Martin15,Martin16}. The absorption of photons passing through the IGM of the cosmic web has been used to constrain the properties of the IGM and to shed light on the physics and nature of reionization (e.g.; see the review by \citealp{Becker15}). The cosmic web is currently used to significantly improve the photometric redshift accuracy of surveys (e.g.; \citealp{Aragon-calvo15}). The structure, properties, and evolution of the cosmic web contain a wealth of information about the initial matter distribution in the universe with valuable cosmological implications (e.g.; see \citealp{Wang16} and the references therein).

Therefore, it is of great importance to characterise and describe the cosmic web of galaxies. However, the multi-scale nature of the cosmic web, its complexity and connectivity, and the lack of a fully objective method in identifying its major components make such studies challenging. Nonetheless, several methods have been developed to quantify and extract the components of the cosmic web in both simulations and observational data (e.g.; see \citealp{Cautun14} for a review). Some of these methods are designed to specifically extract certain components of the web, for example, only filaments \citep{Pimbblet05,Stoica05,Stoica10,Novikov06,Sousbie08,Gonzalez10,Bond10,Smith12,Tempel14}, and some are able to simultaneously break the cosmic web into its major components (e.g.; \citealp{Colberg07,Hahn07,Aragon-Calvo07,Forero-romero09,Aragon-Calvo10b,Jasche10,Sousbie11,Hoffman12,Falck12,Wang12,Leclercq15,Snedden15}). We particularly mention those that take the multi-scale nature of the cosmic web into account, such as the Multi-scale Morphology Filter (MMF) algorithm (\citealp{Aragon-Calvo07}; also see \citealp{Cautun13}). 

These methods have been mostly applied to simulations and some observational data sets with quite interesting results. For example, trends between the dependence of spin, shape, size, and other properties of halos and galaxies on the cosmic web and orientation of filaments and walls are found in simulations \citep{Altay06,Aragon-Calvo07b,Hahn07,Hahn07b,Zhang09,Codis12,Trowland13,Libeskind13,Dubois14,
Welker15,Kang15,Chen15,Chen16,Gonzalez16} and observations \citep{Kashikawa92,Navarro04,Lee07,Paz08,Jones10,Tempel13a,Tempel13b,Zhang13}, generally in support of the Tidal Torque Theory \citep{Peebles69,White84,Codis15} as our comprehension of the origin of the spin of galaxies (also see \citealp{Kiessling15,Joachimi15} for reviews).

Of great interest is the quenching of galaxies in the cosmic web. Generally, two major quenching mechanisms are proposed, the ``environmental quenching'' and ``mass quenching''. The later is thought to be associated with e.g., AGN and stellar feedback (e.g., \citealp{Fabian12,Hopkins14}). The environmental quenching processes such as ram pressure stripping (e.g., \citealp{Gunn72,Abadi99}), galaxy-galaxy interactions and harassment (e.g., \citealp{Farouki81,Merritt83,Moore98}), and strangulation (e.g., \citealp{Larson80,Balogh00}) act in medium to high density environments, with different quenching timescales that seem to depend on galaxy properties as well. Mass quenching has been mostly attributed to central galaxies, whereas environmental quenching is primarily linked to satellites (e.g., \citealp{Peng12,Kovac14}). Moreover, environmental and mass quenching processes seem to suppress star-formation activity independent of each other (e.g., \citealp{Peng10,Quadri12}), although this has been questioned recently. For example, \cite{Darvish16} showed that environmental quenching is more efficient for more massive galaxies and mass quenching is more efficient in denser environments. Interestingly, \cite{Aragon-calvo16} recently showed that the stripping of the filamentary web around galaxies is responsible for star-formation quenching, without the need for feedback processes. 

However, it is still not fully clear whether the environmental effects on galaxy quenching are a local phenomenon or also act on global large-scale cosmic web environments as well. For example, the ``galactic conformity'' --- the observation that satellites are more likely to be quenched around quiescent centrals than star-forming ones --- has been found on both small and larger megaparsec scales \citep{Weinmann06,Kauffmann13,Hartley15,Hearin15,Kawinwanichakij16,Hatfield16,Berti16}, suggesting the role of the large-scale gravitational tidal field on galaxy properties. Moreover, several observations have found that the star-formation activity and other galaxy properties depend on the large-scale cosmic web \citep{Porter08,Fadda08,Biviano11,Ricciardelli14,Darvish14,Darvish15b,Guo15,Chen15b,Alpaslan16,Pandey16}, whereas others have seen no or at best a weak dependence between the properties of galaxies and the global cosmic web environments \citep{Filho15,Eardley15,Penny15,Alpaslan15,Brouwer16,Beygu16,Alonso16,Vulcani16}.  

Nonetheless, the majority of these cosmic web studies are limited to numerical simulations or large spectroscopic surveys in the local universe such as SDSS and GAMA (e.g.; \citealp{Tempel14,Alpaslan14}), mainly due to the completeness, selection function, and projection effect issues involved in observations. Using spectroscopic samples has the benefit of constructing the density field in three dimensions (3D) which suffers less from projection effects. Moreover, establishing the vectorial properties of the cosmic web, for example, the direction of filaments in 3D is possible. However, redshift-space distortions such as the finger-of-god effect should be carefully taken into account so that components of the cosmic web would not be misclassified (e.g., filaments vs. finger-of-god elongated clusters). 

To extend the cosmic web studies to higher redshifts, one could alternatively use photometric information in two-dimensional (2D) redshift slices, as long as the uncertainties in the photometric redshifts are not too large. The information contained in 3D vectorial properties of the cosmic web is usually lost in 2D analyses. However, the scalar quantities such as star-formation rate and stellar mass of galaxies in the cosmic web, on average, and in a statistical sense, are still measurable in 2D projections. The higher redshift studies of the cosmic web are particularly important as its components are not fully evolved and gravitationally merged yet, and much information regarding the properties of galaxies and dark matter haloes, that would otherwise get lost due to the non-linear-interaction regime, is still maintained (e.g.; see \citealp{Jones10,Cautun14}). This sets the need for contiguous large-volume surveys at higher redshifts, with negligible cosmic variance, that are equipped with very accurate photometric redshifts to high-$z$. The COSMOS field survey \citep{Scoville07} is ideal for such cosmic web studies to higher redshifts. 

In pilot studies to target the cosmic web, \cite{Darvish14,Darvish15b} used the 2D version of the MMF algorithm and applied it to potential large-scale structures in the COSMOS at $z$ $\sim$ 0.83 and 0.53. The $z$ $\sim$ 0.83 structure clearly showed a filament linking several clusters and groups and was traced by the distribution of H$\alpha$ emitters \citep{Sobral11,Darvish14}. Further studies of the structure showed that although stellar mass, SFR, and the main-sequence of star-forming galaxies are invariant to the cosmic web, the fraction of H$\alpha$ emitters is enhanced in filaments, likely due to galaxy-galaxy interactions. The other potential filament at $z$ $\sim$ 0.53 was spectroscopically confirmed, and the spectroscopic analysis also showed that although many properties of star-forming galaxies, such as stellar-to-dynamical mass ratio and ionization parameters are independent of their cosmic web environment, gas-phase metallicities are slightly higher in filaments relative to the field and electron densities are significantly lower \citep{Darvish15b}. These are properties shared with star-forming galaxies found in merging clusters, potentially suggesting a connection \citep{Sobral15}.  

These single-structure studies show the potential role of the cosmic web on galaxy evolution. However, small sample size is one of the major issues in these studies. The robustness of our cosmic web detection algorithm in revealing the large-scale cosmic web, the need for a large, homogeneously-selected sample of galaxies located in different regions of the web and extended to higher-$z$, and the limited number of studies that consider the explicit role of the comic web on galaxy evolution, motivate us to extend our analysis to a reliably large sample of galaxies in the whole COSMOS field to $z$ $\sim$ 1.2. Therefore, the focus of this paper is to provide a catalog of density field of galaxies, cosmic web components, and their galactic content over a large and reliable redshift range to the community. We also investigate the star-formation activity of central and satellite galaxies in the global cosmic web environments.               

The format of this paper is as follows. In Section \ref{data}, we briefly review the data. Section \ref{method} outlines the methods used to determine the density field, the comic web extraction, galaxy classification, and the SFR and stellar mass estimation for our sample. In Section \ref{science}, we present the main results, discuss, and compare them with the literature. A summary of this work is given in Section \ref{sum}.

Throughout this work, we assume a flat $\Lambda$CDM cosmology with $H_{0}$=70 kms$^{-1}$ Mpc$^{-1}$, $\Omega_{m}$=0.3 and $\Omega_{\Lambda}$=0.7. All magnitudes are in the AB system and star-formation rates and stellar masses are based on a Chabrier \citep{Chabrier03} initial mass function (IMF). 

\section{Data and Sample Selection} \label{data}

In this work, we use the $\sim$ 1.8 deg$^{2}$ COSMOS field \citep{Scoville07,Capak07} which is ideal for the large scale structure studies at $z\gtrsim$ 0.1, with minimal cosmic variance and a wealth of ancillary data. Using the \cite{Moster11} recipe, the cosmic variance even for the most massive galaxies (log($M/M_{\odot}$) $>$ 11) in this field is only $\sim$ 15-10 \% at $z\sim$ 0.1-3. 

Here, we use the latest COSMOS2015 photometric redshift (photo-$z$) catalog \citep{Laigle16} in the Ultra-VISTA-DR2 region \citep{McCracken12,Ilbert13}. This comprises ground- and space-based photometric data in more than 30 bands (Section \ref{photo-z}). We select objects that are flagged as galaxies, located in the range 149.33 $<\alpha_{2000}$(deg) $<$ 150.8 and 1.6 $<\delta_{2000}$(deg) $<$ 2.83, and are in the redshift range 0.1 $< z <$ 1.2. following our discussion in Section \ref{photo-z}, we limit our study to 0.1 $< z <$ 1.2 to guarantee a reliable density field and cosmic web estimation using very accurate photo-$z$s ($\sigma_{\Delta z/(1+z)}$ $\lesssim$ 0.01). 

\begin{figure} 
 \begin{center}
    \includegraphics[width=3.5in]{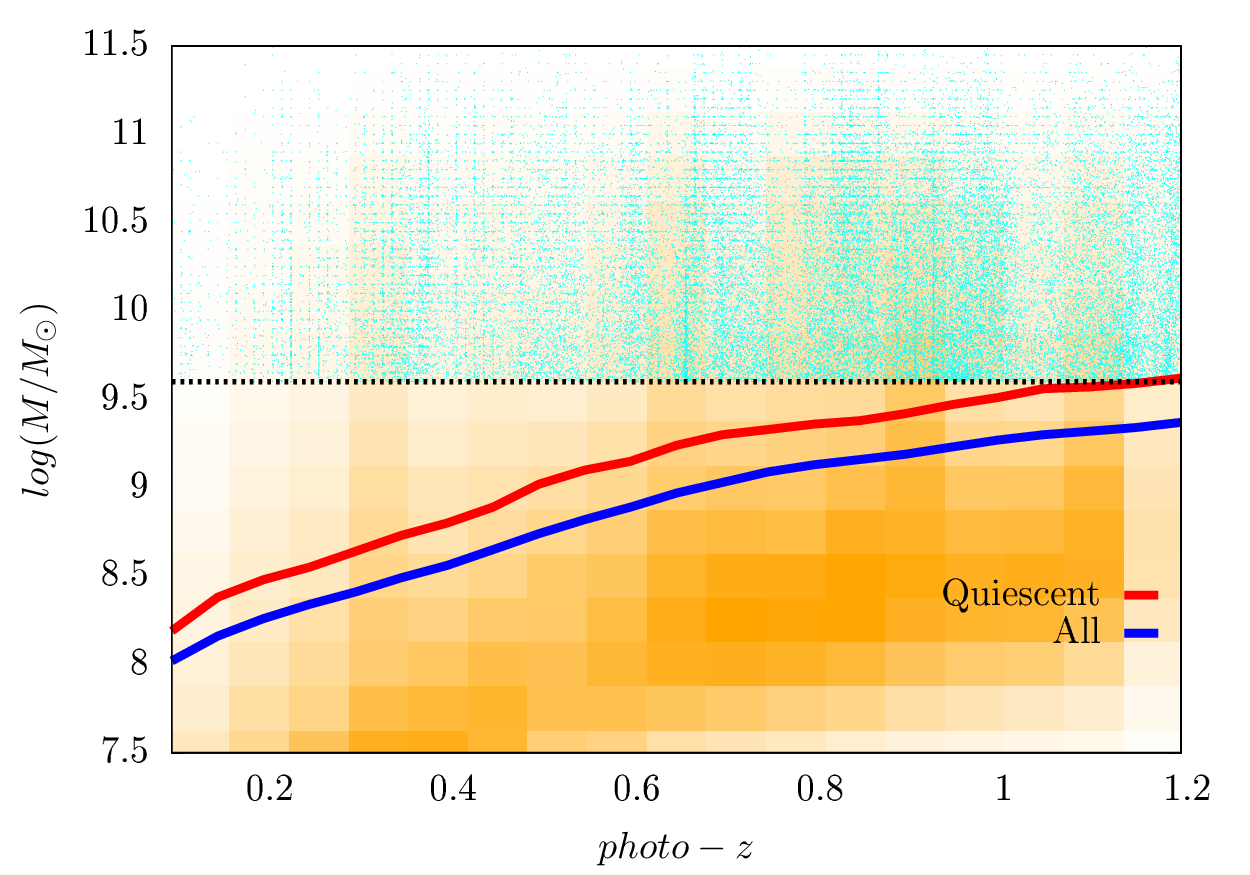}
     \caption{Stellar mass of galaxies as a function of redshift. The blue and red lines show the estimated stellar mass completeness limit for all the galaxies (star-forming and quiescent) and quiescent galaxies only, respectively. All galaxies that are more massive than the mass completeness limit of quiescent galaxies for the highest redshift of this study at $z$=1.2 are selected (log($M/M_{\odot}$) $\geqslant$ 9.6). This is similar to a volume-limited selection. Black dashed line shows this limit and cyan points show our sample galaxies.}
\label{fig:massz}
 \end{center}
\end{figure}

In addition to the aforementioned criteria, we apply a cut based on the stellar mass completeness of the survey (Section \ref{msfrtype}). All galaxies more massive than the mass completeness limit of the highest redshift of this study at $z$=1.2 are selected (log($M/M_{\odot}$) $\geqslant$ 9.6; Section \ref{msfrtype}). This is equivalent to a volume limited sample. We use this sample to estimate the density field (Section \ref{density}), to extract the cosmic web components (Section \ref{web}), and to conduct the analysis in Section \ref{science}. Figure \ref{fig:massz} shows the mass completeness limit and the galaxies selected in this study.

For the analysis here, we only rely on galaxies that are not close to the edge of the field and large masked areas, as the density values and cosmic web assignment for galaxies close to these regions are not reliable. The total number of galaxies before (and after) discarding those near the edge and masked regions is 45421 (38865), respectively. We flag galaxies located near the edge or masked areas in Table 1.

\section{Methods} \label{method}

\subsection{Photo-z Accuracy} \label{photo-z}

In this study, we use the photo-$z$ of galaxies to construct the density field and extract the cosmic web components. Using photometric redshifts automatically suppresses the redshift-space distortions such as the finger-of-god effect. However, large photometric redshift uncertainties would erode and smooth out the real structures in the density field, especially in densest regions. 

A number of studies have shown that using photo-$z$s with typical uncertainties of $\sigma_{\Delta z/(1+z)}$ $\lesssim$ 0.01 can still fairly accurately construct the density field (e.g., \citealp{Cooper05,Malavasi16}), with more optimistic studies such as \cite{Lai16} showing that even larger uncertainties can still reveal the general environmentally-driven trends. Therefore, reliable and accurate photometric redshift measurements are of crucial importance. 

\begin{figure}
 \begin{center}
    \includegraphics[width=3.5in]{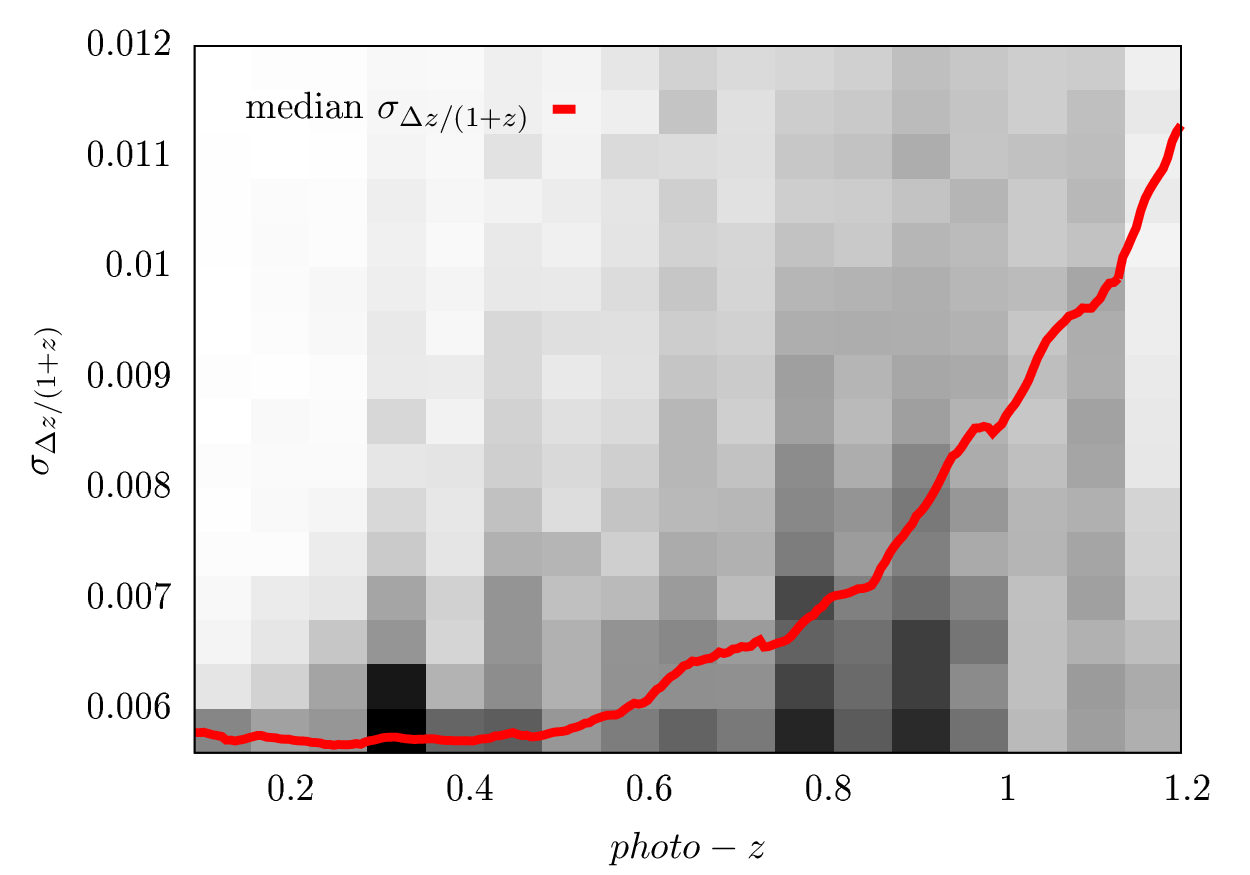}
     \caption{Photo-$z$ uncertainties, $\sigma_{\Delta z/(1+z)}$, as a function of redshift for our sample galaxies. Red line shows the median photo-$z$ uncertainties. We see that the median $\sigma_{\Delta z/(1+z)}$ $\lesssim$ 0.01 out to $z\sim$ 1.2, small enough for reliable construction of the density field and the cosmic web in this work to $z\sim$ 1.2.}
\label{fig:zerr}
 \end{center}
\end{figure}

Here, we use the photometric redshifts from the COSMOS2015 catalog \citep{Laigle16}, which are estimated using over 30 bands from near-UV to far-IR wavelengths. A comparison with the zCOSMOS bright spectroscopic redshift sample \citep{Lilly09} to $z\sim$ 1 shows that photo-$z$ accuracy is $\sigma_{\Delta z/(1+z_{s})}$ $\sim$ 0.007, with a catastrophic failure fraction of only $\sim$ 0.5 \% \citep{Laigle16}. Figure \ref{fig:zerr} shows the photo-$z$ uncertainties, $\sigma_{\Delta z/(1+z)}$, as a function of redshift for our sample, along with the median photo-$z$ uncertainties (red line). Median uncertainties are estimated within $\pm$ 0.2 redshift intervals at each redshift. We clearly see that median $\sigma_{\Delta z/(1+z)}$ $\lesssim$ 0.01 out to $z\sim$ 1.2, consistent with the photo-$z$ vs. spectroscopic redshift comparison, and small enough for reliable construction of the density field and the cosmic web to $z\sim$ 1.2.

\subsection{Stellar Mass, SFR, and Galaxy Classification} \label{msfrtype}

SFRs and stellar masses are based on \cite{Laigle16}, using a SED template fitting procedure similar to that of \cite{Ilbert15} using UV to mid-IR data. The templates were generated using BC03 \citep{Bruzual03}, assuming a Chabrier IMF, two metallicities, a combination of exponentially declining and delayed star formation histories, and two extinction curves. Nebular emission line contributions were considered using an empirical relation between the UV and emission line fluxes \citep{Ilbert09}. The typical stellar mass and SED-based SFR uncertainties for our sample galaxies to $z\sim$ 1.2 are $\Delta M\sim$ 0.05 dex and $\Delta SFR_{SED}\sim$ 0.1 dex, respectively.

To check the reliability of the SED-based SFRs, we compare them with those based on the bolometric IR luminosity for galaxies with a detection in one of $Herschel$ PACS (100 and 160 $\mu$m) and $Herschel$ SPIRE (250, 350, and 500 $\mu$m) bands \citep{Lee13,Lee15a}. This comprises $\sim$ 10\% of the total galaxies. We find a good agreement between the two SFR indicators, with no significant bias and a median absolute deviation of $\sim$ 0.25 dex between them.

The 3$\sigma$ magnitude limit of the survey ($K_{s}$=24; \citealp{Laigle16}) results in a variable stellar mass completeness limit at different redshifts. Using the empirical method originally developed by \cite{Pozzetti10} (see also \citealp{Ilbert13,Darvish15a}), we estimate the stellar mass completeness limit by associating a limiting mass to each galaxy at each redshift. This limiting mass corresponds to the stellar mass that the galaxy would require to have to be detected at its redshift, if its apparent magnitude were equal to the magnitude limit of $K_{s}$=24:
\begin{equation} \label{eq1}
log(M_{limit}/M_{\odot})=log(M/M_{\odot})+0.4(K_{s}-24)
\end{equation}
At each redshift, we define the mass completeness limit as the stellar mass for which 90\% of galaxies have their limiting mass below it. The stellar mass completeness limit also depends on the galaxy type and is higher for quiescent systems. In this study, we rely on the mass completeness limit for quiescent galaxies. 

We separate quiescent galaxies from star-forming systems using rest-frame NUV$-r^{+}$ versus $r^{+}-$J color$-$color plot, with quiescent galaxies satisfying the color selection NUV$-r^{+}$ $>$ 3.1 and NUV$-r^{+}$ $>$ 3($r^{+}-$J)+1 \citep{Ilbert13}. Figure \ref{fig:massz} shows the estimated mass completeness limit for all the galaxies and the quiescent systems, and our sample of galaxies selected for this study (Section \ref{data}).
  
\subsection{Density Field Construction} \label{density}

The density field construction is fully described in \cite{Darvish15a}. Here, we provide a summary and some revisions to the original method. We estimate the density field for a series of overlapping redshift slices ($z$-slice) with variable widths to $z$=1.2. As suggested by \cite{Malavasi16}, $z$-slice widths are selected to be within $\pm$ 1.5 $\sigma_{\Delta z/(1+z)}$ from the center of each redshift (this is slightly different than the widths originally defined in \citealp{Darvish15a}). Then, for each $z$-slice, we associate a weight to each galaxy by measuring the percentage of the photo-$z$ probability distribution function (PDF) of the galaxy that lies within the boundaries of each $z$-slice. This shows the likelihood of a galaxy belonging to that $z$-slice. At each $z$-slice, all galaxies that have weights $\geqslant$ 10 \% are selected for density estimation. The incorporation of the weights tends to significantly diminish the projection effect due to the uncertainties in the photo-$z$s. Figures \ref{fig:map360} (a), \ref{fig:map530} (a), and \ref{fig:map980} (a) show the galaxies selected for density estimation for $z$-slices centered at $z$=0.360, 0.530, and 0.980, respectively. The size of each point is scaled with the weight of that galaxy.  

Through extensive simulations, \cite{Darvish15a} showed that adaptive kernel smoothing (also see \citealp{Scoville07b,Scoville13}) and Voronoi tessellation perform better in constructing the density field compared to other estimators such as the nearest neighbor and Delaunay triangulation. Here, we use the weighted adaptive kernel smoothing (where weights are the assigned galaxy weights explained before) using a 2D Gaussian kernel whose width adaptively changes over the field according to the local density of galaxies. The global smoothing width is selected to be 0.5 Mpc which corresponds to the typical virial radius for X-ray groups and clusters in the COSMOS field \citep{Finoguenov07,George11}. Figures \ref{fig:map360} (b), \ref{fig:map530} (b), and \ref{fig:map980} (b) show the estimated density maps for $z$-slices at $z$=0.360, 0.530, and 0.980, respectively. 

\begin{figure}
 \begin{center}
    \includegraphics[width=3.5in]{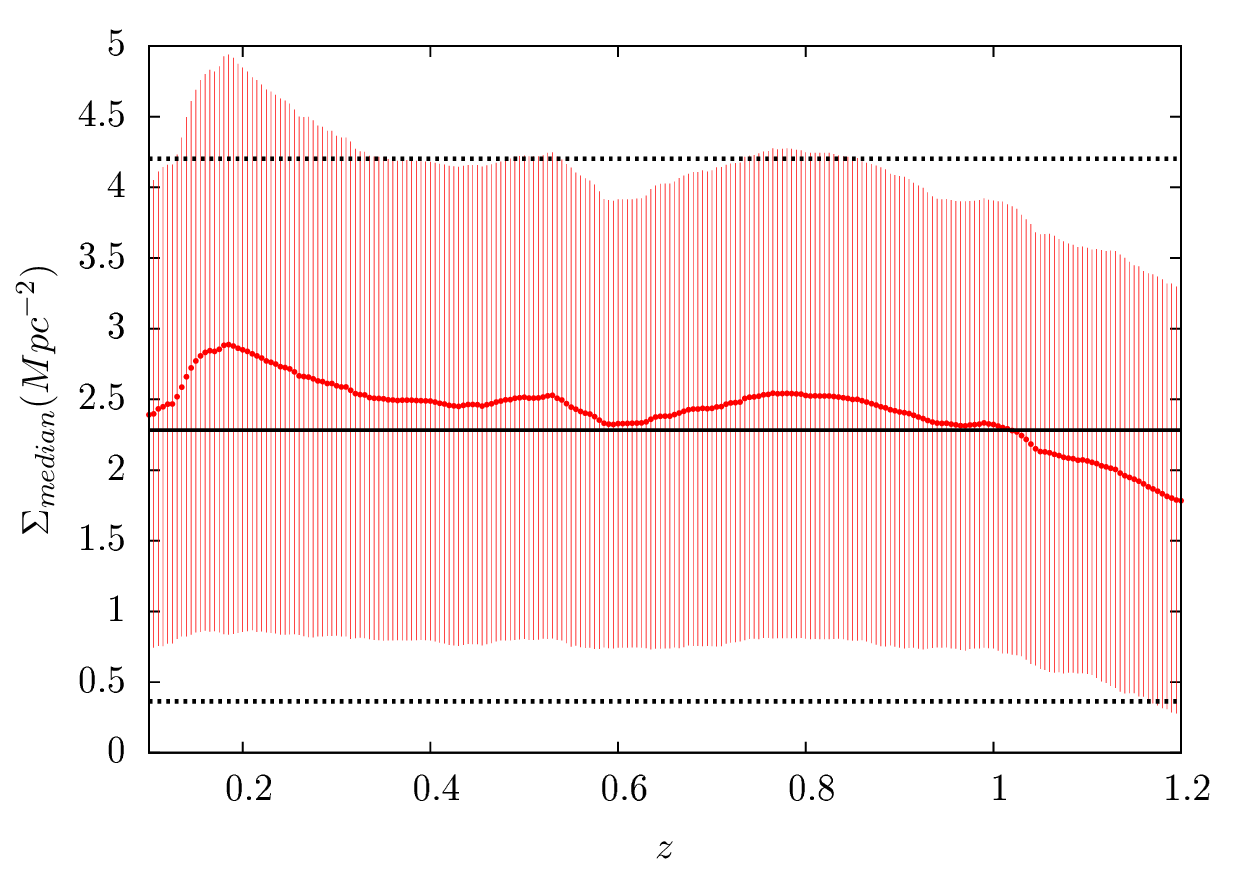}
     \caption{Red symbols show the estimated median density as a function of redshift for our sample galaxies. To minimize the cosmic variance, median densities are estimated within $\pm$ 0.2 redshift intervals at each redshift. The uncertainties are 1.4826 $\times$ the median absolute deviation from the median values. The black solid line shows the estimated median density over the whole field (0.1 $< z <$ 1.2) and the black dashed lines show its upper and lower uncertainties. We find that within the uncertainties, the median values do not change much with redshift. This emphasizes a volume-limited-like sample selection to avoid any unrealistic underestimation of the density values at higher $z$ as a result of missing fainter, less-massive galaxies.}
\label{fig:meden}
 \end{center}
\end{figure}

In constructing the density field, we use our sample of galaxies which is similar to a volume limited sample. This avoids any unrealistic underestimation of the density estimates at higher redshift as less massive, fainter galaxies would be missed at those redshifts. Figure \ref{fig:meden} shows the median density as a function of redshift, along with the median density estimated from the whole field. We see that within the uncertainties, the median values do not change much with redshift. To minimize the cosmic variance, median densities are estimated within $\pm$ 0.2 redshift intervals at each redshift. The uncertainties are 1.4826 $\times$ the median absolute deviation from the median values. 

We finally interpolate our sample galaxies to the estimated density field using their angular position and photo-$z$ PDF. The $z$-slice for each galaxy is the one at which its weight maximizes. Table 1 lists all our sample galaxies with their estimated density values. The overdensity defined as the density with respect to the median density at each redshift ($\Sigma/\Sigma_{median}$) is also given.

\subsection{Cosmic Web Extraction} \label{web}

\subsubsection{The Method} \label{webmtd}

We extract the components of the cosmic web (filaments and clusters) in the density field using the 2D version of the Multi-scale Morphology Filter algorithm \citep{Aragon-Calvo07,Darvish14}. In this method, we associate a filament and a cluster signal to each point (values between 0 and 1) in the density field based on the resemblance of the local geometry of that point to a filament or a cluster. The local geometry of each point is calculated based on the signs and ratio of eigenvalues of the Hessian matrix $H(\bf r)$ which is the second-order derivative of the density field $\Sigma(\bf r)$:
\begin{equation}
H(\bf r)=
\begin{bmatrix}
\nabla_{xx}\Sigma(\bf r)& \nabla_{xy}\Sigma(\bf r)\\
\nabla_{yx}\Sigma(\bf r)& \nabla_{yy}\Sigma(\bf r)
\end{bmatrix}
\end{equation}
where $\nabla_{ij}$s denote the second-order derivatives in the $i$ and $j$ directions. 

Since structures in the density field (filaments and clusters) have different physical sizes, we build a scale-independent structure map by smoothing the surface density field over a range of physical scales and eventually selecting the greatest cluster and filament signal among all the various signal values at different physical scales. In practice, we use a 2D Gaussian smoothing function with physical scales $L$=0.25, 0.50, 0.75, 1.00, 1.50, and 2.00 Mpc. Therefore, the components of the Hessian matrix at scale $L$ are:  
\begin{equation}
\begin{aligned}
H_L(\bf{r}) &= \Sigma\otimes\nabla_{ij}G_L \\
                &= \int d\textbf{r}^{\prime}\Sigma(\textbf{r}^{\prime})\frac{(x_i-x_i^{\prime})(x_j-x_j^{\prime})-\delta_{ij}L^2}{L^4}G_L(\textbf{r}^{\prime},\textbf{r})
\end{aligned}
\end{equation}
where $x_{1}$,$x_{2}$=$x$,$y$,  $x_{1}^{\prime}$,$x_{2}^{\prime}$=$x^{\prime}$,$y^{\prime}$, $\delta_{ij}$ is the Kronecker delta, and $G_L(\textbf{r}^{\prime},\textbf{r})$ is our 2D Gaussian smoothing function at scale $L$:
\begin{equation}
G_L(\textbf{r}^{\prime},\textbf{r})=\frac{1}{2\pi L^{2}}exp(-\frac{\vert \textbf{r}^{\prime}-\textbf{r}\vert}{2L^2})
\end{equation}

\begin{figure}
 \begin{center}
    \includegraphics[width=3.5in]{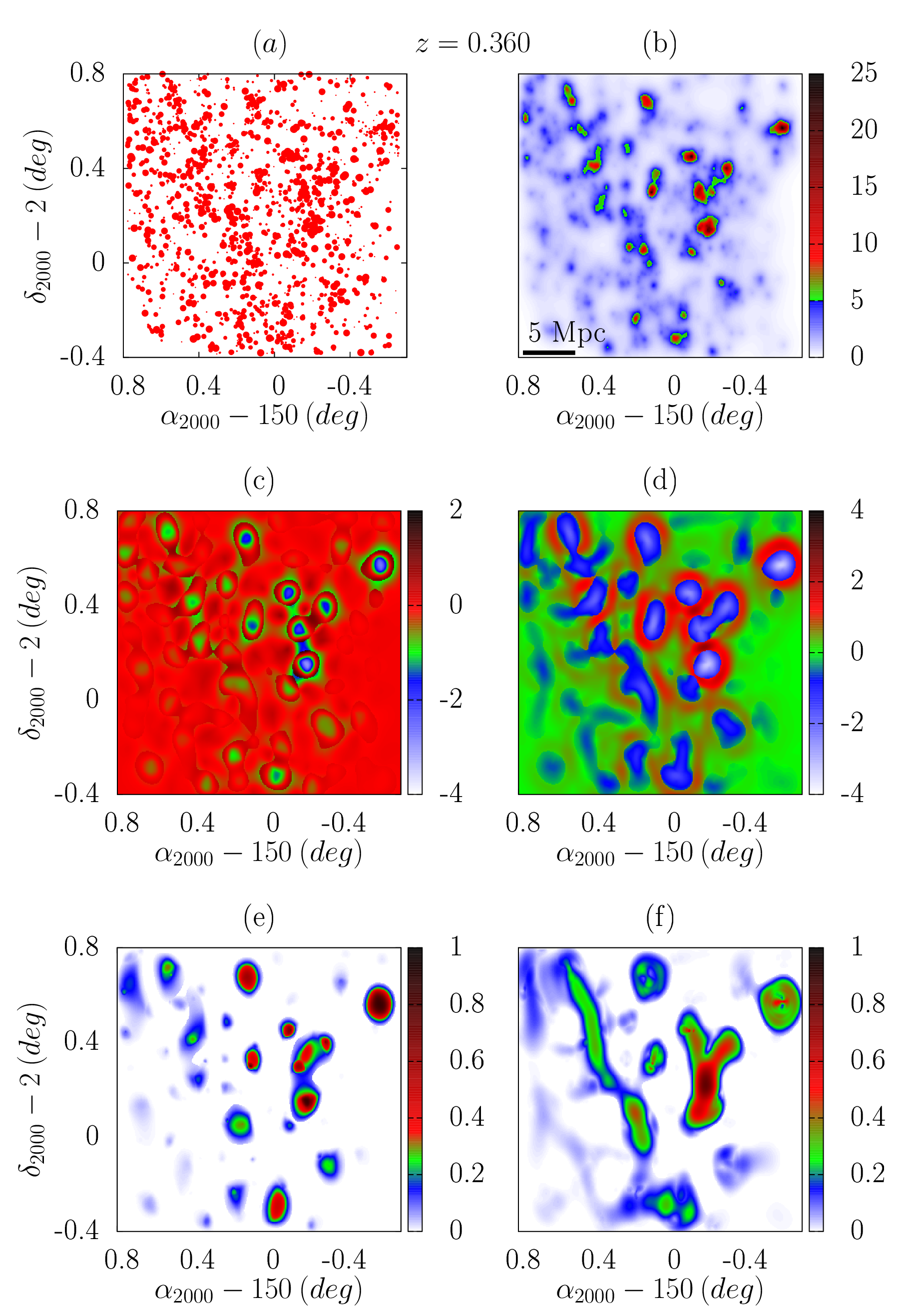}
     \caption{(a) Galaxies selected for density estimation for a $z$-slice centered at $z$=0.360. The size of each point is proportional to the weight assigned to each galaxy (Section \ref{density}). (b) Density field estimated using the weighted adaptive kernel smoothing estimator (Section \ref{density}) for the selected galaxies. (c) $\lambda_{1}$ eigenvalue map of the Hessian matrix evaluated at the physical scale $L$=1.00 Mpc. (d) $\lambda_{2}$ eigenvalue map of the Hessian matrix evaluated at the physical scale $L$=1.00 Mpc, assuming $\vert \lambda_{2}\vert$ $\geqslant$ $\vert \lambda_{1}\vert$. (e) final cluster signal map for the $z$-slice at $z$=0.360 after taking the multi-scale nature of the cosmic web into account. Note that for an ideal cluster we have $\vert \lambda_{1}\vert$ $\approx$ $\vert \lambda_{2}\vert$. (f) final filament signal map for the $z$-slice at $z$=0.360 after taking the multi-scale nature of the cosmic web into account. Note that for an ideal filament we have $\vert \lambda_{1}\vert$ $\ll$ $\vert \lambda_{2}\vert$.}
\label{fig:map360}
 \end{center}
\end{figure} 

\begin{figure}
 \begin{center}
    \includegraphics[width=3.5in]{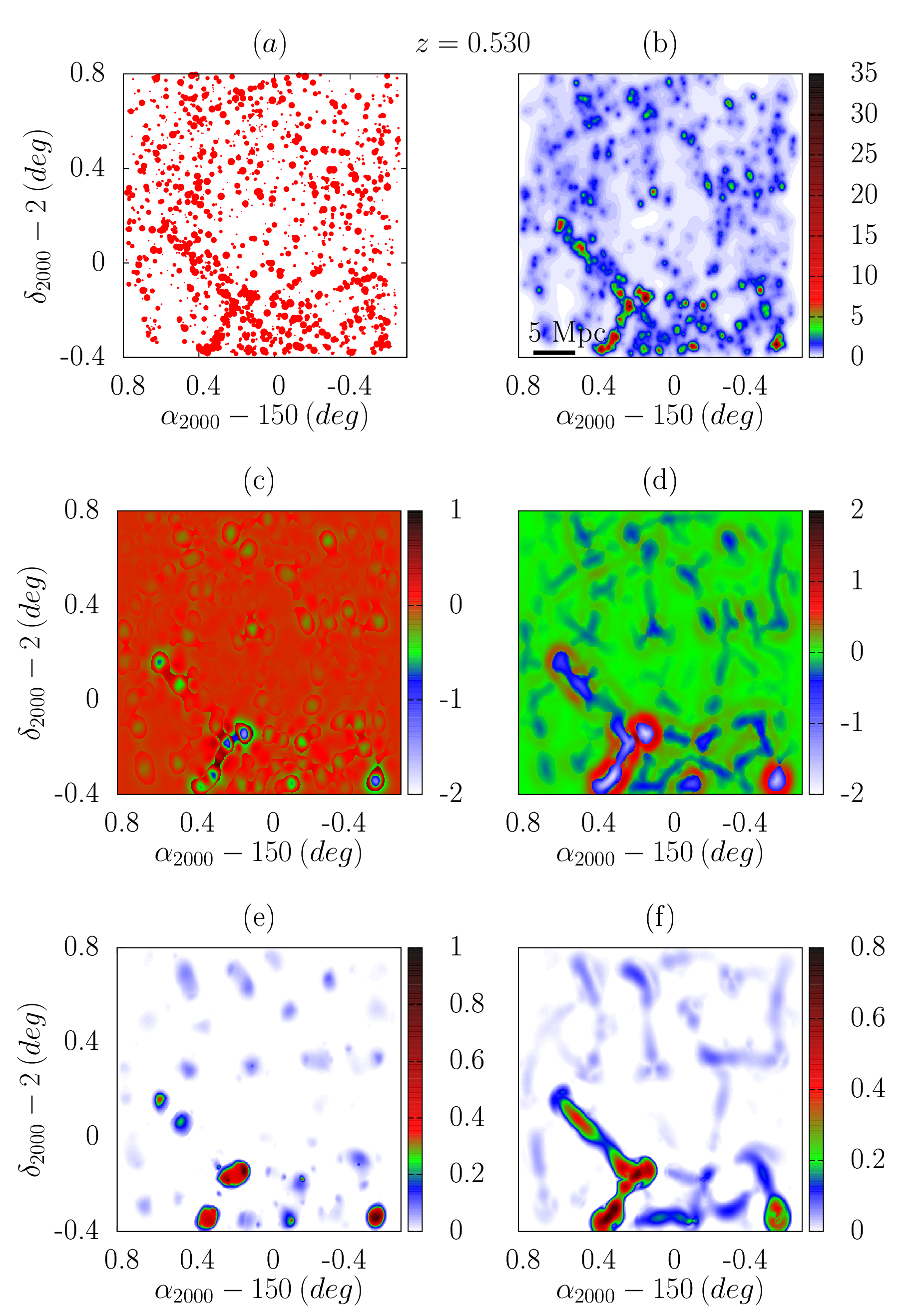}
     \caption{Similar to Figure \ref{fig:map360} but for a $z$-slice centered at $z$=0.530.}
\label{fig:map530}
 \end{center}
\end{figure} 

\begin{figure}
 \begin{center}
    \includegraphics[width=3.5in]{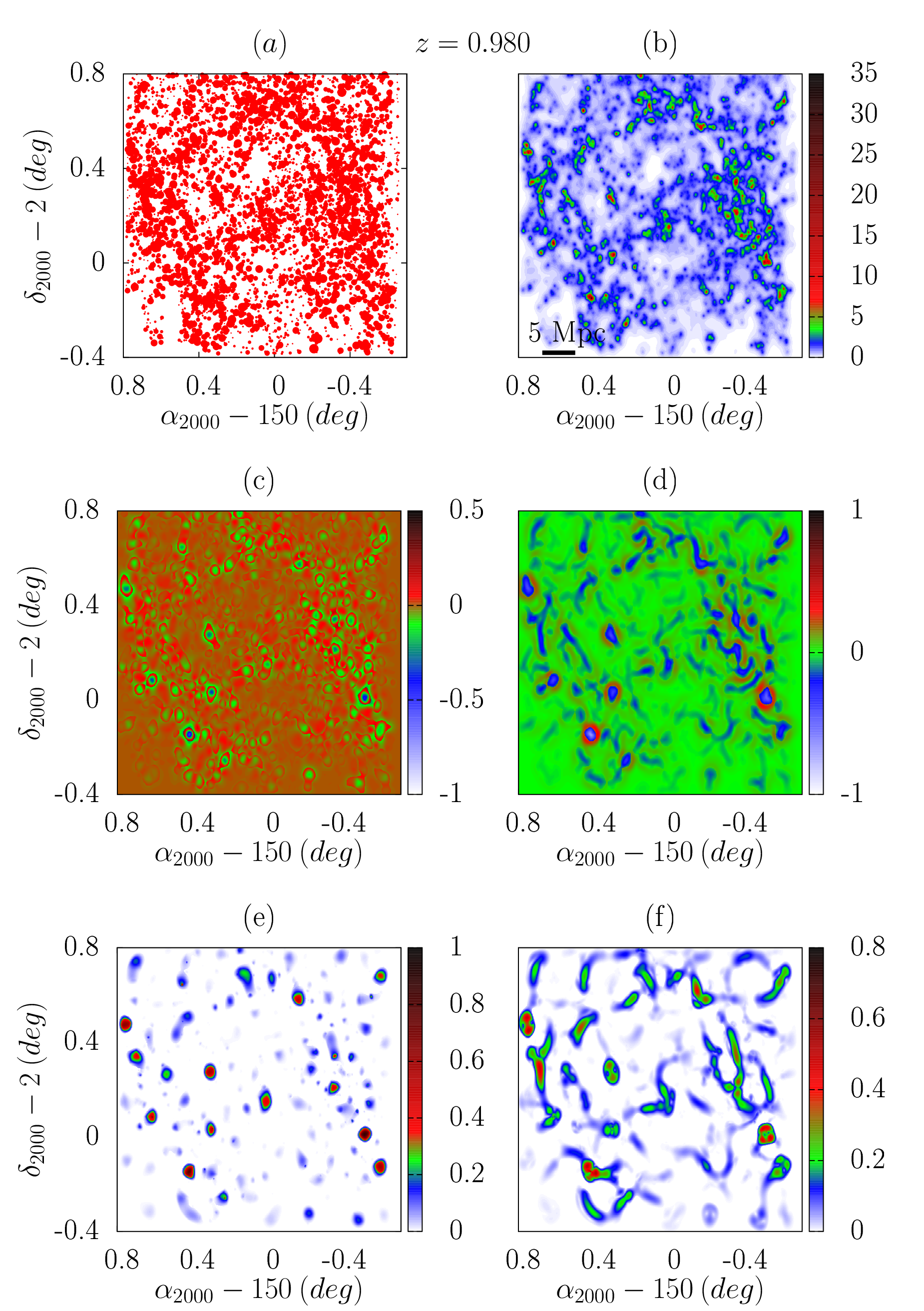}
     \caption{Similar to Figure \ref{fig:map360} but for a $z$-slice centered at $z$=0.980.}
\label{fig:map980}
 \end{center}
\end{figure}

The eigenvalues of the Hessian matrix at each point are a measure of the rate of change of the density field gradient in their corresponding eigenvector direction. Therefore, if the local geometry of a point resembles a cluster, one expects the local eigenvalues to be negative and their ratio close to one. For a filament, one expects the eigenvalue corresponding to the eigenvector perpendicular to the filament to be negative and the ratio of the smaller eigenvalue (in the direction of the filament where the rate of change of density values is small) to the larger one to be close to zero. Figures \ref{fig:map360} (c) and (d), \ref{fig:map530} (c) and (d), and \ref{fig:map980} (c) and (d) show the eigenvalue maps ($\lambda_{1}$ and $\lambda_{2}$) at the physical scale $L$=1.00 Mpc for $z$-slices at $z$=0.360, 0.630, and 0.980, respectively. Given these, if $\lambda_{1}$ and $\lambda_{2}$ are the eigenvalues and $\vert \lambda_{2}\vert$ $\geqslant$ $\vert \lambda_{1}\vert$, we define the morphology mask $\varepsilon$ for clusters and filaments at each point in the density field as:
\begin{equation}
\begin{aligned}
\varepsilon_{cluster} &= 0\ if\ \lambda_1 > 0\ or\ \lambda_2 > 0;\ 1\ otherwise \\ 
\varepsilon_{filament} &= 0\ if\ \lambda_2 > 0;\ 1\ otherwise
\end{aligned}
\end{equation}

For those points in the density field that pass the above conditions ($\varepsilon$=1), we quantify the degree of resemblance to a cluster or a filament by defining the function:
\begin{equation}
\begin{aligned}
D_{cluster} &=\frac{\vert \lambda_{1}\vert}{\vert \lambda_{2}\vert} \\
D_{filament} &=1-\frac{\vert \lambda_{1}\vert}{\vert \lambda_{2}\vert} 
\end{aligned}
\end{equation}

Note that for an ideal cluster ($\vert \lambda_{1}\vert$ $\approx$ $\vert \lambda_{2}\vert$), we have $D_{cluster}$ $\approx$ 1 and $D_{filament}$ $\approx$ 0, whereas for an ideal filament ($\vert \lambda_{1}\vert$ $\ll$ $\vert \lambda_{2}\vert$), we get $D_{cluster}$ $\approx$ 0 and $D_{filament}$ $\approx$ 1.
  
Using the already-defined function $D$, we define the following function $M$ for clusters and filaments \citep{Frangi98}:
\begin{equation}
\begin{aligned}
M_{cluster} &=exp(-\frac{D_{filament}}{2\beta^2}) \\ 
M_{filament} &=exp(-\frac{D_{cluster}}{2\beta^2})
\end{aligned}
\end{equation}
where $\beta$ controls the aggressiveness of feature selection. Here we choose $\beta$=0.5 as a typical value \citep{Frangi98,Aragon-Calvo07}. Note again that for clusters, $D_{filament}$ is small and therefore $M_{cluster}$ is large, whereas for filaments,  $D_{cluster}$ is small and hence $M_{filament}$ is large. 

Another important piece of information that we can use to enhance the detection of structures is that features of our interest (clusters and filaments) are more pronounced in the density field than the overall background distribution. If we do not take this into account, random background fluctuations may result in unrealistic features. For the background, the magnitude of second-order derivatives (and hence eigenvalues) is small due to the lack of contrast. Therefore, we use the norm of the Hessian matrix by defining the function:
\begin{equation}
I=1-exp(-\frac{Norm^{2}}{2c^2}) 
\end{equation}  
where $Norm$=$\sqrt{\lambda_{1}^2+\lambda_{2}^2}$ and $c$ controls the sensitivity of this function. Here we use $c$=0.5$\times$maximum($Norm$) at each $z$-slice \citep{Frangi98}.

Finally, at each scale $L$, the signal map is defined as:
\begin{equation}
S_L=\varepsilon\otimes M\otimes I
\end{equation}
and eventually, every pixel of the final signal map $S$ gets the maximum of all the corresponding pixels at different physical scales. That is:
\begin{equation}
S=max(S_L)
\end{equation}   
Figures \ref{fig:map360} (e) and (f), \ref{fig:map530} (e) and (f), and \ref{fig:map980} (e) and (f) show the final cluster ($S_{cluster}$) and filament ($S_{filament}$) signal maps for the $z$-slices at $z$=0.360, 0.630, and 0.980, respectively. Filament and cluster signal values for our sample galaxies are given in Table 1. 

\begin{figure}
 \begin{center}
    \includegraphics[width=3.5in,height=8in]{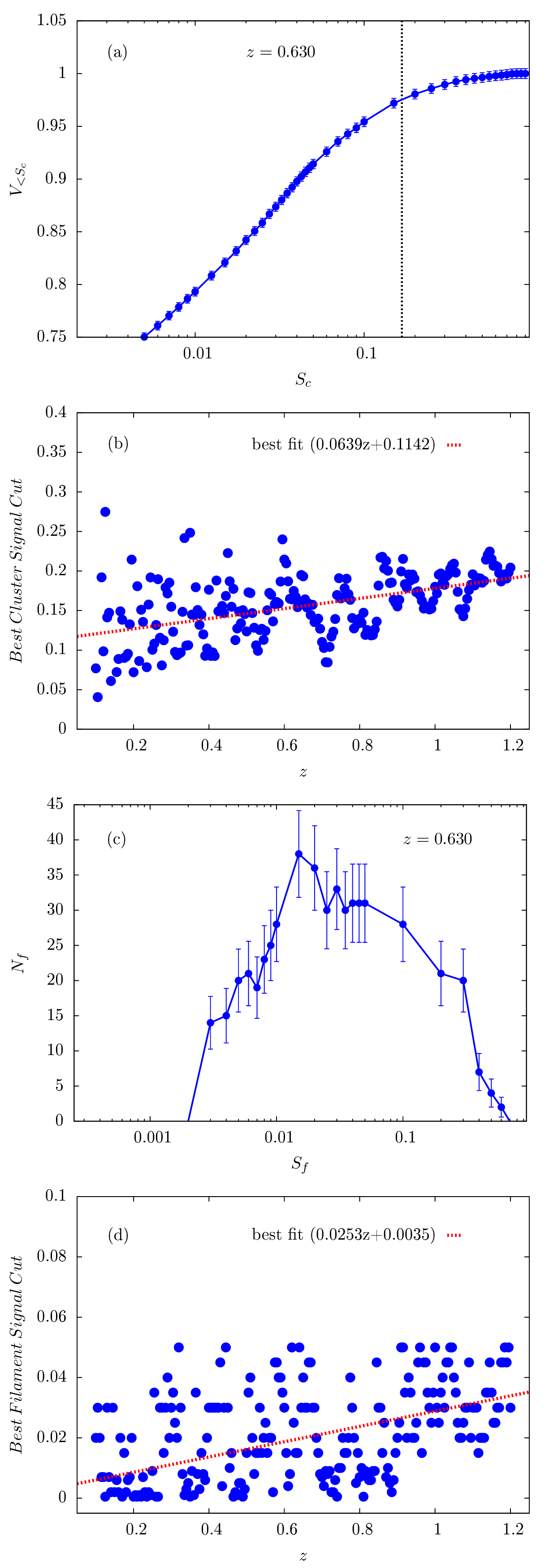}
     \caption{(a) Fraction of volume occupied by below the cluster signal ($V_{< S_c}$) as a function of the cluster signal ($S_{c}$) for the $z$-slice at $z$=0.630. We see a monotonically increasing function which can be described by a two power-law behaviour. The cluster signal that corresponds to the intersection of the two power-law functions (shown as the black dashed line) is selected as the best cluster signal cut. (b) Best cluster signal cut as a function of redshift. It varies from slice to slice and is affected by the cosmic variance. We fit a linear function to the best signal cut at different redshifts and use the fitted line, shown with a red dashed line, for selection of cluster galaxies. (c) Number of individual filaments ($N_{f}$) as a function of filament signal cut ($S_{f}$) for the $z$-slice at $z$=0.630. We use the $S_{f}$ corresponding to the peak of $N_{f}$ as the best filament signal cut (Section \ref{webslctn}). (d) Best filament signal cut as a function of redshift. We fit a linear function to the best signal cut at different redshifts and use the fitted line, shown with a red dashed line, for selection of filament galaxies.}
\label{fig:filclrparam}
 \end{center}
\end{figure}

\subsubsection{Filament, Cluster, and Field Selection} \label{webslctn}

For each $z$-slice, there is a filament and a cluster signal assigned to each point which is a value between 0 and 1. Choosing a very high signal value results in a small sample size and disregards possible real structures with small signal values, whereas a very small signal is prone to contamination from unreal features and noise in the density field. Therefore, an appropriate, trade-off signal cut for filaments and clusters should be chosen.

\begin{table*}
\begin{center}
{\scriptsize
{Table 1}
\begin{tabular}{lcccccccccccc}
\hline
\noalign{\smallskip}
$\alpha_{2000}$ & $\delta_{2000}$ & photo-z & density & overdensity & cluster & filament & cosmic web & group ID & number of & galaxy & flag\tablenotemark{a}\\
 (deg) & (deg) & & (Mpc$^{-2}$) & & signal & signal & environment & & group members & type &\\
\hline
150.041038 & 1.679104 & 0.2200 & 2.1721 & 0.7777 & 0.000411 & 0.236093 & filament & 74 & 3 & central & 0\\
149.468224 & 1.660186 & 0.6036 & 3.3760 & 1.4498 & 0.050772 & 0.062457 & filament & 1656 & 5 & central & 0\\
149.854923 & 1.661894 & 0.2611 & 0.3327 & 0.1248 & 0.000004 & 0.002267 & field & -99 & -99 & singleton & 0\\
149.849106 & 1.660836 & 0.7437 & 1.9569 & 0.7773 & 0.133678 & 0.218654 & filament & 2602 & 14 & satellite & 0\\
149.570287 & 1.660729 & 0.6966 & 1.0058 & 0.4132 & 0.001817 & 0.010613 & field & -99 & -99 & singleton & 0\\
149.431513 & 1.660398 & 0.9849 & 0.2869 & 0.1234 & 0.000000 & 0.001720 & field & 5302 & 2 & central & 0\\
149.734774 & 1.660589 & 0.7550 & 0.7674 & 0.3030 & 0.000413 & 0.000150 & field & -99 & -99 & singleton & 0\\
150.261428 & 1.660046 & 0.9069 & 7.3397 & 3.0500 & 0.161472 & 0.155610 & field & 4043 & 5 & satellite & 0\\
149.596887 & 1.660786 & 0.8669 & 0.4738 & 0.1910 & 0.000309 & 0.006520 & field & 3845 & 3 & central & 0\\
149.942127 & 1.660848 & 1.0543 & 2.1166 & 0.9947 & 0.057414 & 0.114930 & filament & -99 & -99 & singleton & 0\\
149.778184 & 1.660923 & 0.9624 & 1.1430 & 0.4926 & 0.085444 & 0.056621 & field & -99 & -99 & singleton & 0\\
149.466702 & 1.660667 & 0.6359 & 4.9136 & 2.0826 & 0.318805 & 0.445941 & filament & 1830 & 3 & central & 0\\
149.434744 & 1.660370 & 0.5339 & 29.4506 & 11.7433 & 0.810066 & 0.312929 & cluster & 1177 & 30 & satellite & 0\\
150.363551 & 1.661725 & 0.9211 & 1.1555 & 0.4865 & 0.022983 & 0.097828 & filament & -99 & -99 & singleton & 0\\
149.633649 & 1.660999 & 0.6832 & 2.2096 & 0.9084 & 0.001875 & 0.008153 & field & 2237 & 2 & central & 0\\
150.332107 & 1.661299 & 1.0417 & 0.7554 & 0.3459 & 0.010454 & 0.027575 & field & 5763 & 2 & satellite & 0\\
149.489826 & 1.660462 & 0.7497 & 1.4808 & 0.5870 & 0.004759 & 0.020155 & field & -99 & -99 & singleton & 0\\
149.420016 & 1.661311 & 0.9977 & 0.5942 & 0.2559 & 0.012427 & 0.015011 & field & 5302 & 2 & satellite & 0\\
150.033681 & 1.661263 & 0.8750 & 3.2449 & 1.3192 & 0.068370 & 0.040883 & field & 3890 & 2 & satellite & 0\\
149.822392 & 1.661513 & 0.9007 & 0.2286 & 0.0948 & 0.000000 & 0.000670 & field & -99 & -99 & singleton & 0\\
150.309442 & 1.661536 & 1.1818 & 1.8980 & 1.0359 & 0.077422 & 0.058708 & field & -99 & -99 & singleton & 0\\
149.807899 & 1.660575 & 0.5954 & 2.2641 & 0.9751 & 0.177001 & 0.313080 & filament & 1559 & 10 & satellite & 0\\
149.458808 & 1.660600 & 0.6356 & 5.9750 & 2.5324 & 0.395283 & 0.498088 & filament & 1830 & 3 & satellite & 0\\
150.161337 & 1.661601 & 0.7401 & 6.6049 & 2.6260 & 0.187206 & 0.146879 & cluster & 2501 & 23 & satellite & 0\\
149.802478 & 1.661148 & 0.6603 & 2.3846 & 0.9965 & 0.050753 & 0.129584 & filament & 1997 & 3 & satellite & 0\\
149.810858 & 1.662660 & 1.1455 & 1.0050 & 0.5161 & 0.037894 & 0.043417 & filament & -99 & -99 & singleton & 0\\
149.421077 & 1.661087 & 0.5080 & 8.6137 & 3.4325 & 0.464037 & 0.186550 & cluster & 1135 & 5 & satellite & 0\\
150.137783 & 1.662048 & 1.1956 & 0.6520 & 0.3646 & 0.003127 & 0.003832 & field & -99 & -99 & singleton & 0\\
149.425513 & 1.662000 & 0.5735 & 2.3355 & 0.9821 & 0.006330 & 0.137193 & filament & 1420 & 3 & satellite & 0\\
150.321295 & 1.662851 & 1.1326 & 2.2188 & 1.1212 & 0.109606 & 0.076917 & field & 6438 & 2 & satellite & 0\\
150.315566 & 1.661740 & 0.5327 & 7.4348 & 2.9646 & 0.385201 & 0.576236 & filament & 1134 & 52 & satellite & 0\\
149.959531 & 1.661993 & 0.3744 & 13.2503 & 5.3112 & 0.234709 & 0.058717 & cluster & 412 & 47 & satellite & 0\\
149.942584 & 1.662862 & 1.0847 & 1.4336 & 0.6892 & 0.024018 & 0.091628 & filament & -99 & -99 & singleton & 0\\
150.264656 & 1.663255 & 0.9347 & 2.4810 & 1.0606 & 0.026360 & 0.073873 & filament & 4516 & 6 & satellite & 0\\
150.078485 & 1.662787 & 0.7150 & 2.5167 & 1.0206 & 0.005093 & 0.002531 & field & 2417 & 5 & satellite & 0\\
149.769251 & 1.663466 & 0.9346 & 1.6804 & 0.7184 & 0.102289 & 0.076040 & field & 4515 & 2 & satellite & 0\\
150.333574 & 1.663059 & 1.1175 & 2.6052 & 1.2829 & 0.094398 & 0.039504 & field & 6147 & 3 & central & 0\\
150.263827 & 1.662800 & 1.1122 & 2.5708 & 1.2563 & 0.030129 & 0.021809 & field & -99 & -99 & singleton & 0\\
149.518530 & 1.662758 & 0.9833 & 1.9457 & 0.8371 & 0.107281 & 0.082587 & field & -99 & -99 & singleton & 0\\
149.831415 & 1.662806 & 0.8337 & 3.6871 & 1.4654 & 0.021558 & 0.014761 & field & 3219 & 6 & satellite & 0\\
\hline
\end{tabular}
\tablecomments{(a) Objects that are close to the edge or masked areas are flagged 1. Otherwise, they are flagged 0.}
\tablecomments{Table 1 is published in its entirety in the electronic edition of the {\it Astrophysical Journal}. A portion is shown here for guidance regarding its form and content.}
}
\end{center}
\end{table*}

Following \cite{Aragon-Calvo07}, we select the appropriate cluster and signal cuts at different redshifts. For clusters, if we plot the fraction of volume occupied by below the cluster signal ($V_{< S_c}$) as a function of the cluster signal ($S_{c}$), we see a monotonically increasing function which can be described by a two power-law behaviour (Figure \ref{fig:filclrparam} (a) for the $z$-slice at $z$=0.630). The cluster signal that corresponds to the intersection of the two power-law functions is selected as the best cluster signal cut. Figure \ref{fig:filclrparam} (a) shows this for one of the $z$-slices. The best cluster signal cut varies from slice to slice and is likely affected by the cosmic variance. Hence, we fit a linear function to the best signal cut at different redshifts and use the fitted line (0.0639$z$+0.1142) for selection (Figure \ref{fig:filclrparam} (b)). The typical best cluster signal cut is in the range $\sim$0.1-0.2.          

For filaments, the number of individual filaments ($N_{f}$) at very small filament signal cut ($S_{f}$) is small because pixels tend to percolate and form large filaments. At very large $S_{f}$ values, $N_{f}$ is also small because only a small fraction of pixels pass the selection cut. Therefore, if we plot $N_{f}$ versus $S_{f}$, it maximizes at some $S_{f}$ value (Figure \ref{fig:filclrparam} (c) for the $z$-slice at $z$=0.630). We use this $S_{f}$ corresponding to the peak of $N_{f}$ as the best filament signal cut. Similar to the cluster selection, we use the fitted line (0.0253$z$+0.0035) results to select the best filament signal cut at different redshifts (Figure \ref{fig:filclrparam} (d)). The typical best filament signal cut is in the range $\sim$0.01-0.04.

At each redshift, all the galaxies that have their cluster signal greater than (or equal to) the best cluster signal cut at that redshift and their cluster signal greater than (or equal to) their filament signal are selected as cluster galaxies. We use the remaining points to impose the filament selection. Among the remaining points, all the galaxies that have their filament signal greater than (or equal to) the best filament signal cut at that redshift and their filament signal greater than (or equal to) their cluster signal are selected as filament galaxies. Eventually, the final remaining points that do not satisfy both filament and cluster selections are chosen as the field. Table 1 contains the cosmic web environment of our sample galaxies.

\begin{figure}
 \begin{center}
    \includegraphics[width=3.5in]{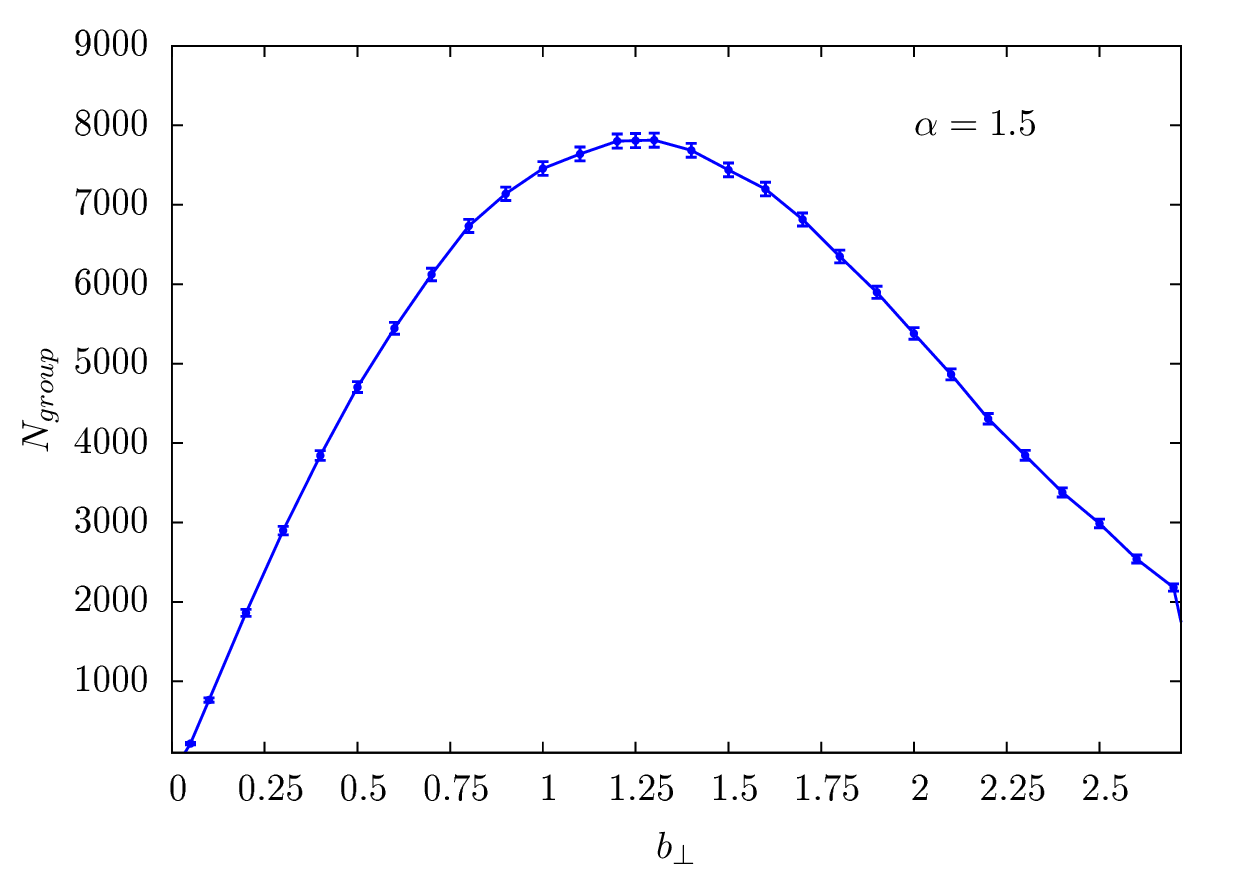}
     \caption{Number of groups $N_{group}$ as a function of the projected linking length in units of the median projected intergalaxy separation ($b_{\perp}$) for a fixed $\alpha$=1.5 (Section \ref{centsat}). The number of groups for small $b_{\perp}$ values is low because only a small number of galaxies are too close to link together. On the other hand, for large values of $b_{\perp}$, $N_{group}$ is also small because many galaxies link together to form percolated structures. By fixing $\alpha$=1.5, the number of selected groups maximizes at $b_{\perp}$ $\sim$ 1.3 and we use it as a suitable, trade-off linking length for our group selection.}
\label{fig:grprm}
 \end{center}
\end{figure}

\subsection{Central, Satellite, and Singleton Classification} \label{centsat}

We select a sample of galaxy groups and use it to observationally classify central and satellite galaxies in our data set. We select the most massive galaxy in each group as a central and the rest as satellites. Galaxies that are not associated with any galaxy group (singletons) are either centrals whose satellites, in principle, are too faint to be detected in our volume-limited sample or they are ejected satellites orbiting beyond their halo's virial radius (see e.g.; \citealp{Wetzel14}). We rely on our sample galaxies (our volume-limited-like sample) to identify groups. We use the commonly-used friends-of-friends algorithm (\citealp{Huchra82}; also see \citealp{Duarte14} and the references therein), by linking galaxies whose redshift difference and angular separations are less than some critical values. Two galaxies $i$ and $j$ with redshifts $z_{i}$ and $z_{j}$ and angular separation $\Delta \theta_{ij}$ are linked to each other if they satisfy the following conditions:
\begin{equation}
\begin{aligned}
D_c(z)\Delta \theta_{ij} &\leqslant b_{\perp} n(z)^{-1/2} \\
\vert z_i-z_j\vert &\leqslant \alpha \sigma_{\Delta z/(1+z)}  
\end{aligned}
\end{equation}
where $D_c(z)$ is the comoving distance at $z$ (average redshift of galaxies $i$ and $j$), $n(z)$ is the median number density of galaxies at $z$ (Figure \ref{fig:meden}), $b_{\perp}$ is the projected linking length in units of the median projected intergalaxy separation at $z$, and $\alpha$ is a parameter that controls the line-of-sight linking of galaxies as a function of the typical photo-$z$ uncertainties at $z$. 

The appropriate values for $b_{\perp}$ and $\alpha$ are key in selection of our galaxy groups. Small linking lengths tend to break groups into many subcomponents, whereas large linking lengths tend to percolate and link different groups into a single, larger one. Therefore, depending on the science of interest, trade-off linking lengths will be selected. 

Here we use $b_{\perp}$=1.3 and $\alpha$=1.5. The selection of latter is based on our discussion in Section \ref{photo-z} and \cite{Malavasi16}. Similar to our discussion in Section \ref{webslctn}, the number of groups ($N_{group}$) for small $b_{\perp}$ values is low because only a small number of galaxies are too close to link together. On the other hand, for large values of $b_{\perp}$, $N_{group}$ is also small because many galaxies link together to form percolated structures. By fixing $\alpha$=1.5, the number of selected groups maximizes at $b_{\perp}$ $\sim$ 1.3 and we use it as a suitable, trade-off linking length (see Figure \ref{fig:grprm}). We note that fine-tuning the above-mentioned parameters does not significantly change our results. Table 1 lists the group ID, number of group members, and the galaxy type (satellite, central, singleton) of our sample.

\section{Results and Discussion} \label{science}

\subsection{SF Activity in the Cosmic Web} \label{SFR}

\begin{figure*}
 \begin{center}
    \includegraphics[width=5.8in]{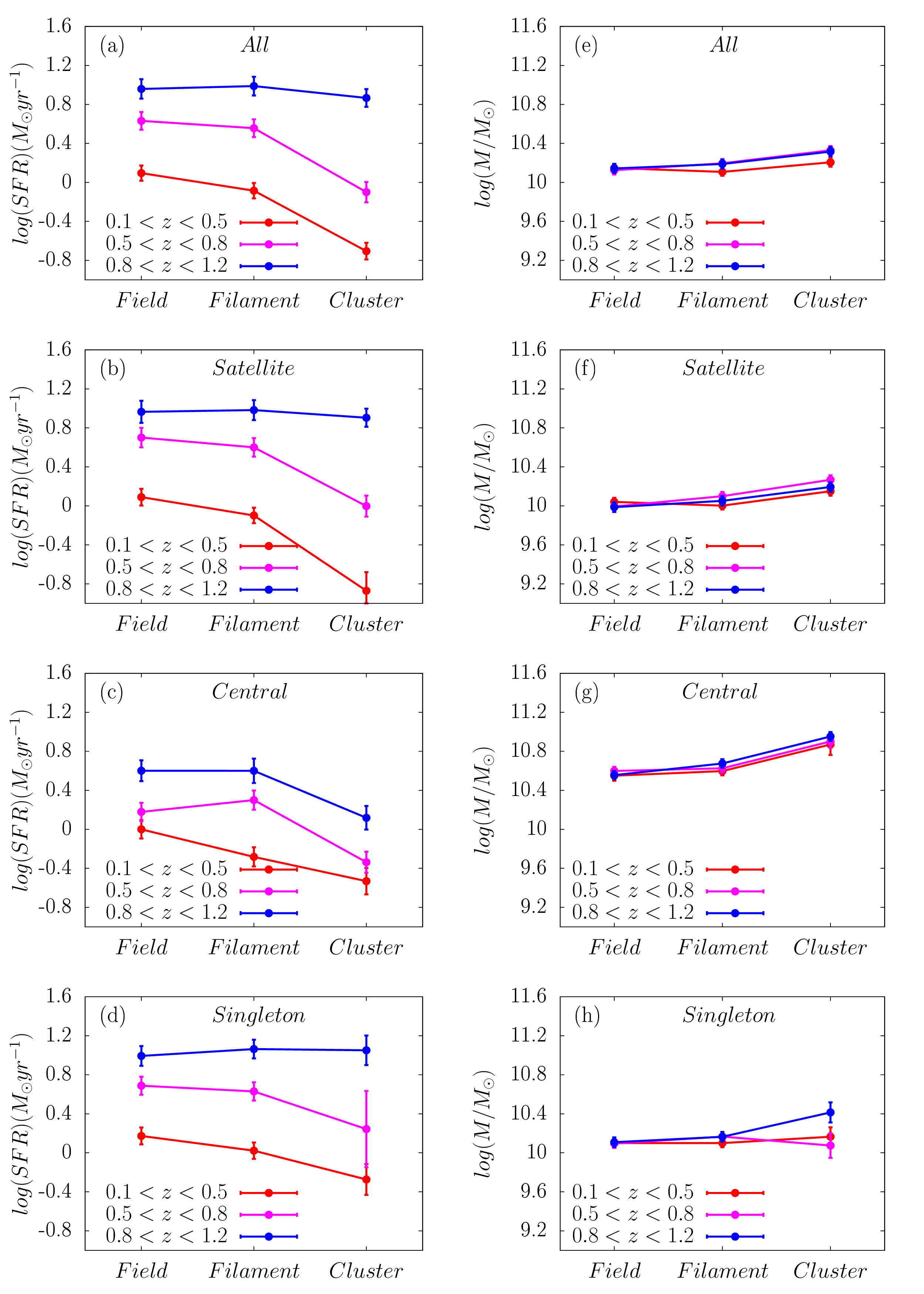}
     \caption{(a) Median SFR for galaxies in different parts of the cosmic web from the field to clusters for different redshift bins. We clearly see a gradual decline in the median SFR  from the field to clusters at $z$ $\lesssim$ 0.8 but at higher redshifts ($z$ $\gtrsim$ 0.8), the trend flattens out. The decline in the median SFR from the field to filaments is not significantly large but the SFR difference between cluster galaxies and those located in other regions of the cosmic web is quite evident at $z$ $\lesssim$ 0.8. (b) Median SFR for satellite galaxies located in different regions of the comic web at different redshifts. These trends are very similar to the overall population of galaxies, indicating the dominance of satellites in determining the general trends. At 0.1 $\leqslant z \leqslant$ 0.5, the median SFR of satellites decreases by $\sim$ 1 dex as one goes from the field to clusters. (c) Median SFR of central galaxies in the comic web at different redshifts. In the whole redshift range considered here, the median SFR of central galaxies decreases by $\sim$ 0.5 dex from the field to clusters. (d) Median SFR of singleton galaxies in the comic web at different redshifts. Singletons at $z$ $\lesssim$ 0.8 show similar trends to centrals, whereas at $z$ $\gtrsim$ 0.8, their trend resembles that of satellites. (e) to (h) Median stellar mass for all, satellite, central, and singleton galaxies in the cosmic web, respectively. Within the uncertainties, we see almost no change or a slight increase in some cases ($\sim$ 0.2-0.3 dex in maximum) in the median stellar mass of galaxies from the field to clusters. Therefore, stellar mass differences in different parts of the cosmic web cannot much explain these trends or make them even stronger.}
\label{fig:sfrmassall}
 \end{center}
\end{figure*}

Figure \ref{fig:sfrmassall} (a) shows the median SFR for galaxies in different parts of the cosmic web from the field to clusters for different redshift bins. Error bars are estimated using the bootstrap resampling, added in quadrature to typical observational uncertainties and uncertainties due to the cosmic variance. We clearly see a gradual decline in the median SFR  from the field to clusters at $z$ $\lesssim$ 0.8 but at higher redshifts ($z$ $\gtrsim$ 0.8), the trend flattens out. The decline in the median SFR from the field to filaments is not significantly large but the SFR difference between cluster galaxies and those located in other regions of the cosmic web is quite evident at $z$ $\lesssim$ 0.8.  

We further investigate this relation for satellite, central, and singleton galaxies as shown in Figure \ref{fig:sfrmassall} (b), (c), and (d). For satellites, the trends are very similar to the overall population of galaxies, indicating the dominance of satellites in determining the general trends. For centrals and at $z$ $\lesssim$ 0.8, we also see a decline in the median SFR from the field to clusters although the decline is not as sharp as that of satellites in the same redshift range. For examples, for satellites at 0.1 $\leqslant z \leqslant$ 0.5, the median SFR decreases by $\sim$ 1 dex as one goes from the field to clusters, whereas centrals show a $\sim$ 0.5 dex decline. This shows that the environmental quenching is mostly due to satellites. Singletons at $z$ $\lesssim$ 0.8 show similar trends to centrals. However, at $z$ $\gtrsim$ 0.8, their trend resembles that of satellites. Note that the difference between the median SFR of filament and field galaxies is not significant within the uncertainties.  

\begin{figure*}
 \begin{center}
    \includegraphics[width=5.8in]{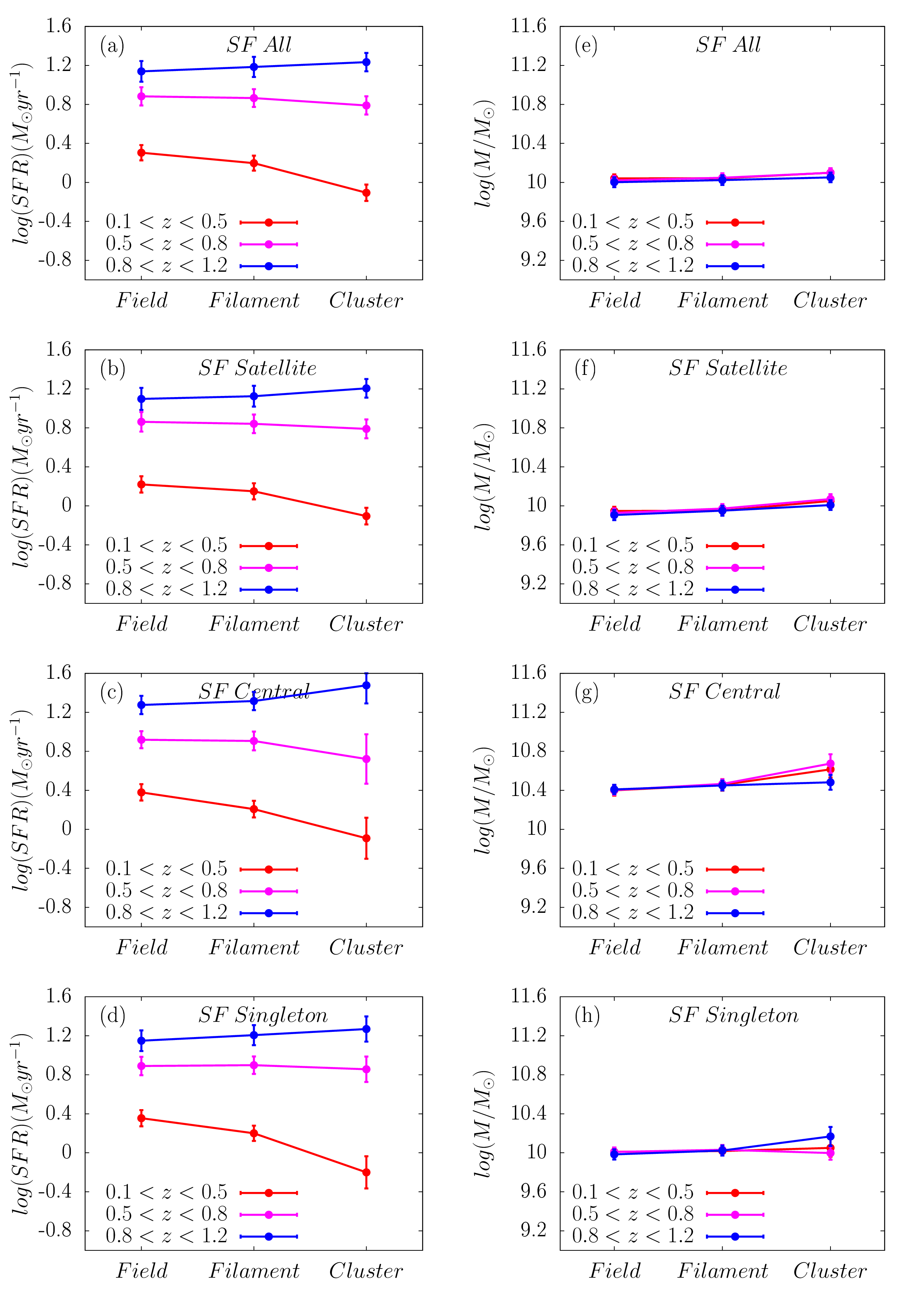}
     \caption{(a) to (d) Median SFR for all star-forming, satellite star-forming, central star-forming, and singleton star-forming galaxies in the cosmic web, respectively. At $z$ $\gtrsim$ 0.5, and within the uncertainties, the median SFR for star-forming centrals, satellites, singletons, and all, do not much depend on the cosmic web. However, at $z$ $\lesssim$ 0.5, all satellite, central, and singleton (and all) star-forming galaxies show a $\sim$ 0.3-0.4 dex decline in their median SFR. These results have implications for the nature of galaxy quenching in the cosmic web and their qualitative timescales (Section \ref{science}). (e) to (h) Median stellar mass for all star-forming, satellite star-forming, central star-forming, and singleton star-forming galaxies in the cosmic web, respectively. Within the uncertainties, we see almost no change or a slight increase in some cases ($\sim$ 0.2-0.3 dex in maximum) in the median stellar mass of star-forming galaxies from the field to clusters. Therefore, stellar mass differences in different parts of the cosmic web cannot much explain the observed cosmic web dependence of the SFRs here or make the trends even stronger.}
\label{fig:sfrmasssf}
 \end{center}
\end{figure*}

\begin{figure*}
 \begin{center}
    \includegraphics[width=6.5in]{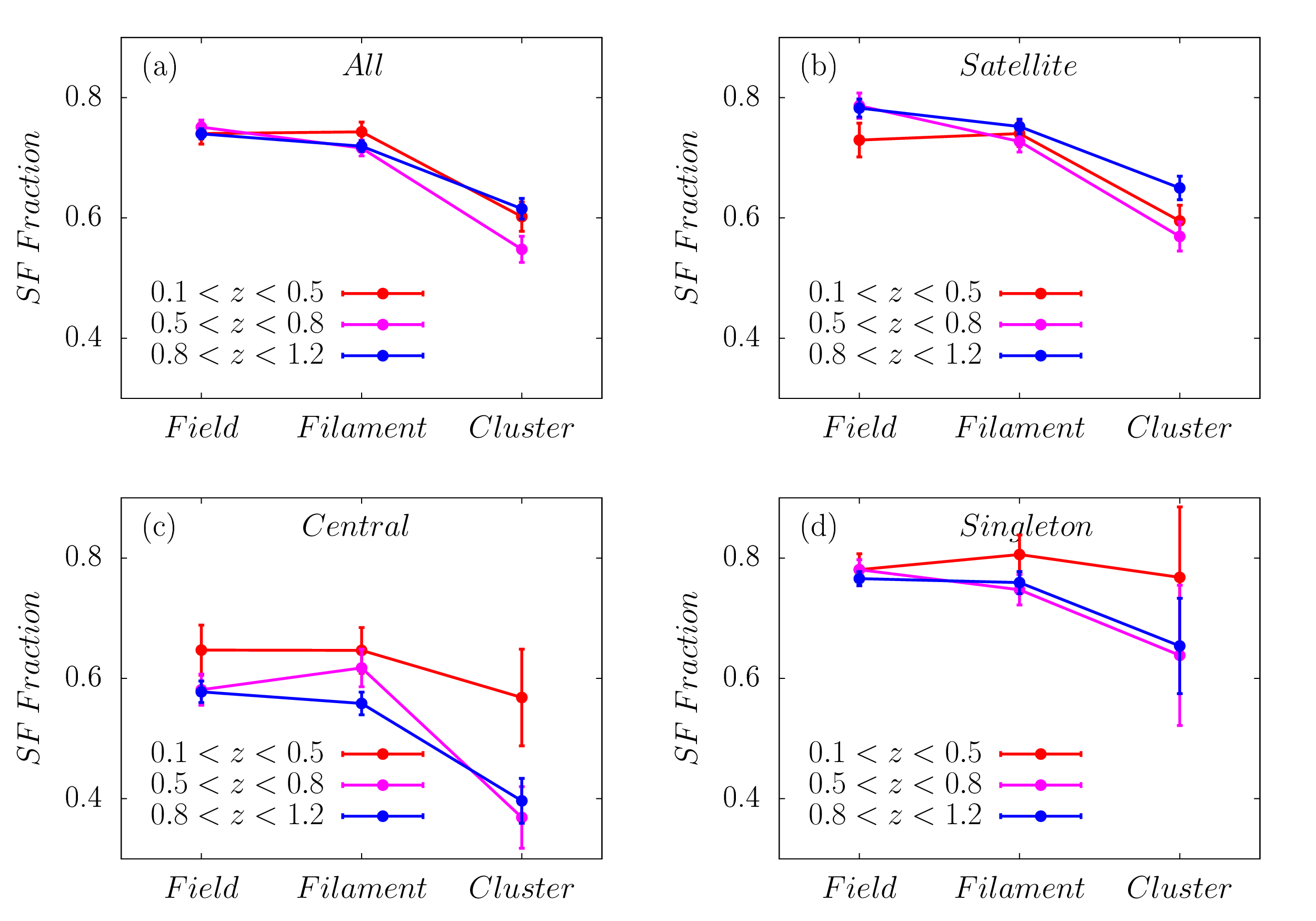}
     \caption{(a) to (d) Star-forming (SF) fraction for all, satellite, central, and singleton galaxies in the cosmic web, respectively. The SF fraction declines from the field to clusters for the overall population of galaxies and for satellite systems at all the redshifts considered in this work, without any significant redshift evolution. At $z$ $\lesssim$ 0.5 and within the uncertainties, the SF fraction is similar in different environments for central and singleton systems and it declines from the field to clusters at $z$ $\gtrsim$ 0.5 for them. Given the results in this figure and Figures \ref{fig:sfrmassall} and \ref{fig:sfrmasssf}, we conclude that most satellites experience a fast quenching mechanism as they fall from the field into clusters through filaments, whereas central and singleton galaxies mostly undergo a fast environmental quenching process at $z$ $\gtrsim$ 0.5 and a slow mechanism at $z$ $\lesssim$ 0.5. Note that the difference between filament and field systems is not significant, within the uncertainties.}
\label{fig:sff}
 \end{center}
\end{figure*}

Interestingly, for central galaxies, we still see an environmental (cosmic web) dependence in the median SFR even at higher redshifts ($z$ $\gtrsim$ 0.8), with a decline of $\sim$ 0.5 dex in the median SFR from the field to clusters. Satellites an singletons do not show any cosmic web dependence at these redshifts.

It is important to consider the role of stellar mass on these rends as well. For example, the decline in  the median SFRs from the field to clusters might be simply due to a different stellar mass distribution in different regions of the cosmic web. We investigate this by estimating the median stellar mass in the cosmic web as shown in Figure \ref{fig:sfrmassall} (e) to (h). Within the uncertainties, we see almost no change or a slight increase in most cases ($\sim$ 0.2-0.3 dex in maximum) in the median stellar mass of galaxies from the field to clusters. Therefore, if we control for stellar mass in different parts of the comic web, these trends do not change or become even stronger. 

For example, for satellite galaxies at 0.8 $\leqslant z \leqslant$ 1.2, the median stellar mass increases by $\sim$ 0.2 dex from the field to clusters (but with large error bars). If we assume that this increase is dominated by star-forming (or quiescent) galaxies, the $\sim$ 0.2 dex stellar mass enhancement from the field to clusters is equivalent to $\sim$ 0.2 dex increase in the SFR, assuming a slope of $\sim$ 1 for the main-sequence of star-forming (or quiescent) galaxies. This means that the median SFR in clusters should in principle decrease by $\sim$ 0.2 dex compared to the field when we control for the stellar mass. This suggests that there might still exist a cosmic web dependence of the median SFR for satellite galaxies even at 0.8 $\leqslant z \leqslant$ 1.2 although the trend is not as strong as those at lower redshifts and the large uncertainties do not allow for a strong conclusion.

The overall decline in the median SFR from the field to clusters can be due to an overall decline in the median SFR of individual galaxies as they fall from the field into clusters through filaments, a decline in the fraction of star-forming galaxies from the field to clusters or both. We investigate this by focusing on star-forming galaxies only. 

Figure \ref{fig:sfrmasssf} (a) to (d) show the median SFR for all star-forming, satellite star-forming, central star-forming, and singleton star-forming galaxies in the cosmic web. We clearly see that at $z$ $\gtrsim$ 0.5, and within the uncertainties, the median SFR for star-forming centrals, satellites, singletons, and all, do not depend much on the cosmic web. However, at $z$ $\lesssim$ 0.5, all satellite central, and singleton (and all) star-forming galaxies show a $\sim$ 0.3-0.4 dex decline in their median SFR. 

Therefore, at $z$ $\gtrsim$ 0.5, the decline in the median SFR from the field to clusters for satellites, centrals, and singletons is mainly due to a change in the fraction of star-forming and quiescent galaxies, whereas at $z$ $\lesssim$ 0.5, the decline is due to a combination of the overall decline in the SFR of individual galaxies and a decline in the fraction of star-forming galaxies from the field to clusters. However, at $z$ $\lesssim$ 0.5 and for central and singleton galaxies, the overall decline in the SFR of individual central and singleton galaxies is the main cause ($\sim$ 0.4-0.5 dex out of $\sim$ 0.5 dex decline can be explained by it), whereas, for satellites, the change in the fraction of star-forming/quiescent satellite galaxies is the main driver of this trend ($\sim$ 0.7 dex out of $\sim$ 1 dex decline can be explained by it). 

Figure \ref{fig:sff} shows this conclusion more clearly. The star-forming (SF) fraction is plotted for all the galaxies, satellites, centrals, and singletons in the cosmic web. We clearly see that the SF fraction for the global trend and satellite systems declines from the field to clusters at all the redshifts considered in this study, without a significant evolution. However, for centrals, the SF fraction declines from the field to clusters only at $z$ $\gtrsim$ 0.5 and is almost unchanged at $z$ $\lesssim$ 0.5 within the uncertainties. Note that within the uncertainties, the change in the SF fraction between field and filament galaxies is not significant. 

\textit{Given these, most satellite galaxies experience a rapid quenching mechanism as they fall from the field into clusters through the channel of filaments, whereas central and singleton galaxies undergo a slow environmental quenching process at $z$ $\lesssim$ 0.5 and a fast mechanism at higher redshifts ($z$ $\gtrsim$ 0.5)}. We note again that controlling the stellar mass does not much affect (or tends to decease) the median SFRs in clusters compared to the field and filaments. Similar results are also found for the relation between the cosmic web and the sSFR of galaxies. 

Using the local overdensity of galaxies as a measure of the environment and a similar dataset, \cite{Darvish16} and \cite{Scoville13} showed that the median SFR for the overall population of galaxies depends on the local environment out to $z$ $\sim$ 1.1-1.2 and is lower in denser regions. We checked this using our current sample of galaxies and found that our results based on the local overdensity of galaxies are in agreement with previous studies \citep{Scoville13,Darvish16}. However, as we showed in Figure \ref{fig:sfrmassall}, the median SFR for the general galaxy population depends on the global cosmic web only to $z$ $\lesssim$ 0.8. This might suggest that the local environment of galaxies is more fundamental than the global cosmic web environment, at least in this redshift range (0.8 $\lesssim$ $z$ $\lesssim$ 1.2). 

\cite{Peng12} in the local universe and \cite{Kovac14} out to $z$ $\sim$ 0.7 showed that satellite quenching is the main driver of the environmental effects. As we showed in Figures \ref{fig:sfrmassall}, \ref{fig:sfrmasssf}, and \ref{fig:sff}, the observed trends for satellites resemble those of the overall galaxy population, suggesting the dominant role of satellite galaxies in shaping the general environmental trends, in agreement with \citep{Peng12,Kovac14}.

For star-forming galaxies, the cosmic web independence of the median SFR at $z$ $\gtrsim$ 0.5 is consistent with \cite{Darvish14,Darvish15b}. \cite{Darvish14} showed that for a $z$ $\sim$ 0.83 LSS, the observed median SFR for H$\alpha$ star-forming galaxies is almost independent of the cosmic web. \cite{Darvish15a} showed that the equivalent width of [OII] line (a measure of the sSFR) as a function of stellar mass is almost the same for filament and field star-forming galaxies at $z$ $\sim$ 0.53. Furthermore, \cite{Erfanianfar16} found that the main-sequence of star-forming galaxies at 0.5 $<$ $z$ $<$ 1.1 is similar for group, filament-like, and field galaxies but at 0.15 $<$ $z$ $<$ 1.1, they found that group galaxies deviate from the main-sequence toward lower SFRs at a fixed stellar mass. These are fully consistent with our results for the star-forming galaxies.

However, as already discussed in \cite{Darvish16}, for star-forming galaxies, there is no consistency on this topic in the literature over different redshifts, as some studies have found an environmental dependence of SFR for star-forming galaxies (e.g.; \citealp{Vulcani10,Vonderlinden10,Patel11,Haines13,Tran15,Erfanianfar16}), whereas others found none/weak (e.g.; \citealp{Patel09,Peng10,Wijesinghe12,Muzzin12,Koyama13a,Ricciardelli14,Lin14,
Darvish14,Vogelsberger14,Cen14,Stroe15,Darvish15b,Vulcani16b}). 

For example, even at 0.1 $\leqslant$ $z$ $<$ 0.5, \cite{Darvish16} found that the median SFR for star-forming galaxies, even at fixed stellar mass, is independent of the local overdensities within the uncertainties. However, using a similar dataset and redshift range in this work (0.1 $\leqslant z \leqslant$ 0.5), we see that satellite, central, and singleton star-forming galaxies show a decline in their median SFR in clusters compared to the field. Part of the difference might be due to this idea that the global cluster membership (and the global halo properties) might be more important than the local overdensities in determining the star-formation activity of star-forming galaxies at lower redshifts. Another possibility is that dividing the galaxies into different overdensity bins results in a small sample size in each bin which might wash out any global environmental dependence of SFR for star-forming galaxies at lower redshifts, particularly when combined with typically large SED-based SFR uncertainties. 

Several studies have seen an enhancement in the fraction of active star-forming galaxies in filaments with respect to clusters and the field, likely due to interaction between galaxies as they fall into denser regions of clusters along the filaments (see e.g.; \citealp{Fadda08,Darvish14}). We do not see a significant enhancement in the SF fraction in filaments for our sample (see Figure \ref{fig:sff}). Part of this discrepancy might be due to the different type of star-forming galaxies with different star formation timescales used. For example, \cite{Darvish14} used H$\alpha$ emitters with much shorter star formation timescales than our current study (which relies on SED template fitting SFRs), and \cite{Fadda08} used starburst galaxies in their study. In other words, if the star formation activity really enhances in filaments in very short timescales, shorter than the SED-based SFRs, we will not be able to see that in this study.         

\cite{Sobral16} showed that H$\alpha$ star-forming galaxies at $z$ $\sim$ 0.4 are significantly dustier than their field counterparts, resulting in apparently lower SFRs in denser environment if dust correction has not been applied in SFR estimation. We note that our SED-based SFR has taken the dust correction into account and in a statistical sense, cannot significantly alter our results here.  

In the local universe, \cite{Poudel16} recently found that at fixed group mass and large-scale luminosity density, central galaxies in filaments have lower sSFR than those outside of filaments. Our results at 0.1 $<$ $z$ $<$ 0.5 for central galaxies (e.g.; Figure \ref{fig:sfrmassall}) show a similar trend in filaments compared to those in the field, consistent with \cite{Poudel16}.
 
By combining the SDSS data with a high-resolution N-body simulation, \cite{Wetzel13} showed that for satellites of log($M/M_{\odot}$) $>$ 9.7 at $z$ $\sim$ 0, SFRs evolve unaffected for 2-4 Gyr after infall into a halo, after which star formation quenches rapidly, with an e-folding timescale of 0.2-0.8 Gyr and shorter quenching timescales for more massive satellites. Recently, \cite{Hahn16} extended this to central galaxies, showing that it takes a total migration time of $\sim$ 4 Gyr from main-sequence to quiescence for log($M/M_{\odot}$)=10.5 central galaxies, $\sim$ 2 Gyr longer than satellites. These studies are qualitatively consistent with our results, suggesting a slower quenching timescale for the majority of centrals at $z$ $\lesssim$ 0.5, possibly due to a slow quenching process such as strangulation. However, for the majority of satellites, a fast quenching mechanism after their infall into their host cluster halos, such as ram pressure stripping, can better explain our results. We highlight that our results also support a fast environmental quenching mechanism for both centrals and satellites at $z$ $\gtrsim$ 0.5. We also note that due to the slow nature of the strangulation mechanism, it makes sense that we only see its effects on the evolution of galaxies only at lower redshifts (i.e.; Figure \ref{fig:sfrmasssf}). 

It is worth noting that despite the large size of the sample used in this study, the overall uncertainties are still large. This sets the need for extremely large-volume surveys in near future such as LSST, Euclid, and WFIRST to substantially tackle this issue.   
   
\section{Summary and Conclusion} \label{sum}

We use a mass complete (log($M/M_{\odot}$) $\geqslant$ 9.6) sample of galaxies in the $\sim$ 2 deg$^{2}$ COSMOS field out to $z$=1.2 to construct the density field from which the comic web of galaxies is extracted. Using the density field Hessian matrix, we disentangle the cosmic web into clusters, filaments and the general field. We provide a catalog of environmental measures such as the local density, cosmic web, and central, satellite, and singleton dichotomy to the community. We investigate the star-formation activity of galaxies in the cosmic web to $z$=1.2 with the following main results:
\begin{enumerate}
\item{Within the uncertainties, we do not find a significant different between the star-formation activity in filaments and the field.}
\item The median SFR of the overall population of galaxies declines in the cosmic web from the field to clusters at $z$ $\lesssim$ 0.8 and flattens out at higher redshifts. Satellite galaxies experience the largest decline of $\sim$ 1 dex especially at lower redshifts, whereas centrals and singletons show a decline of $\sim$0.4-0.5 dex in the same redshift range. 
\item The median SFR of the star-forming galaxies in the cosmic web declines by $\sim$ 0.3-0.4 dex from the field to clusters for satellites, centrals, and singletons at $z$ $\lesssim$ 0.5, and is almost independent of the comic web at higher redshifts. 
\item The star-forming fraction for the overall galaxy population and satellite systems declines from the field to clusters at all the redshifts considered in this work, without any significant redshift evolution. For central galaxies, the SF fraction is almost unchanged in the cosmic web at $z$ $\lesssim$ 0.5 and it declines from the field to clusters at $z$ $\gtrsim$ 0.5. 
\item The decline in the median SFR for satellite galaxies in the cosmic web is mainly due to a decrease in the fraction of satellite star-forming galaxies from the field to clusters, suggesting a rapid quenching mechanism for the majority of satellites in the web. For central galaxies, the slighter decline in the median SFR at $z$ $\lesssim$ 0.5 is mainly attributed to an overall decline in the SFR of individual central galaxies, suggesting a slower quenching process for central systems at $z$ $\lesssim$ 0.5. However, centrals at $z$ $\gtrsim$ 0.5 should also undergo a fast environmental quenching mechanism. 
\end{enumerate}     

This paper is the first one in a series studying the explicit role of the cosmic web on galaxy properties over the past $\sim$ 8 Gyr. In a following paper, we will investigate the dependence of other galaxy properties on the cosmic web and will explain the results in the context of galaxy formation and evolution.     

\section*{acknowledgements}

B.D. acknowledges financial support from NASA through the Astrophysics Data Analysis Program (ADAP), grant number NNX12AE20G. D.S. acknowledges financial support from the Netherlands Organisation for Scientific Research (NWO) through a Veni fellowship and from Lancaster University through an Early Career Internal Grant A100679. 

\bibliographystyle{apj} 
\bibliography{references}

\begin{thebibliography}{}
\expandafter\ifx\csname natexlab\endcsname\relax\def\natexlab#1{#1}\fi

\bibitem[{{Abadi} {et~al.}(1999){Abadi}, {Moore}, \& {Bower}}]{Abadi99}
{Abadi}, M.~G., {Moore}, B., \& {Bower}, R.~G. 1999, \mnras, 308, 947

\bibitem[{{Alonso} {et~al.}(2016){Alonso}, {Hadzhiyska}, \&
  {Strauss}}]{Alonso16}
{Alonso}, D., {Hadzhiyska}, B., \& {Strauss}, M.~A. 2016, \mnras, 460, 256

\bibitem[{{Alpaslan} {et~al.}(2014){Alpaslan}, {Robotham}, {Driver}, {Norberg},
  {Baldry}, {Bauer}, {Bland-Hawthorn}, {Brown}, {Cluver}, {Colless}, {Foster},
  {Hopkins}, {Van Kampen}, {Kelvin}, {Lara-Lopez}, {Liske}, {Lopez-Sanchez},
  {Loveday}, {McNaught-Roberts}, {Merson}, \& {Pimbblet}}]{Alpaslan14}
{Alpaslan}, M., {Robotham}, A.~S.~G., {Driver}, S., {et~al.} 2014, \mnras, 438,
  177

\bibitem[{{Alpaslan} {et~al.}(2015){Alpaslan}, {Driver}, {Robotham},
  {Obreschkow}, {Andrae}, {Cluver}, {Kelvin}, {Lange}, {Owers}, {Taylor},
  {Andrews}, {Bamford}, {Bland-Hawthorn}, {Brough}, {Brown}, {Colless},
  {Davies}, {Eardley}, {Grootes}, {Hopkins}, {Kennedy}, {Liske},
  {Lara-L{\'o}pez}, {L{\'o}pez-S{\'a}nchez}, {Loveday}, {Madore}, {Mahajan},
  {Meyer}, {Moffett}, {Norberg}, {Penny}, {Pimbblet}, {Popescu}, {Seibert}, \&
  {Tuffs}}]{Alpaslan15}
{Alpaslan}, M., {Driver}, S., {Robotham}, A.~S.~G., {et~al.} 2015, \mnras, 451,
  3249

\bibitem[{{Alpaslan} {et~al.}(2016){Alpaslan}, {Grootes}, {Marcum}, {Popescu},
  {Tuffs}, {Bland-Hawthorn}, {Brough}, {Brown}, {Davies}, {Driver}, {Holwerda},
  {Kelvin}, {Lara-L{\'o}pez}, {L{\'o}pez-S{\'a}nchez}, {Loveday}, {Moffett},
  {Taylor}, {Owers}, \& {Robotham}}]{Alpaslan16}
{Alpaslan}, M., {Grootes}, M., {Marcum}, P.~M., {et~al.} 2016, \mnras, 457,
  2287

\bibitem[{{Altay} {et~al.}(2006){Altay}, {Colberg}, \& {Croft}}]{Altay06}
{Altay}, G., {Colberg}, J.~M., \& {Croft}, R.~A.~C. 2006, \mnras, 370, 1422

\bibitem[{{Arag{\'o}n-Calvo} {et~al.}(2007{\natexlab{a}}){Arag{\'o}n-Calvo},
  {Jones}, {van de Weygaert}, \& {van der Hulst}}]{Aragon-Calvo07}
{Arag{\'o}n-Calvo}, M.~A., {Jones}, B.~J.~T., {van de Weygaert}, R., \& {van
  der Hulst}, J.~M. 2007{\natexlab{a}}, \aap, 474, 315

\bibitem[{{Aragon-Calvo} {et~al.}(2016){Aragon-Calvo}, {Neyrinck}, \&
  {Silk}}]{Aragon-calvo16}
{Aragon-Calvo}, M.~A., {Neyrinck}, M.~C., \& {Silk}, J. 2016, ArXiv e-prints,
  arXiv:1607.07881

\bibitem[{{Arag{\'o}n-Calvo} {et~al.}(2010{\natexlab{a}}){Arag{\'o}n-Calvo},
  {Platen}, {van de Weygaert}, \& {Szalay}}]{Aragon-Calvo10b}
{Arag{\'o}n-Calvo}, M.~A., {Platen}, E., {van de Weygaert}, R., \& {Szalay},
  A.~S. 2010{\natexlab{a}}, \apj, 723, 364

\bibitem[{{Arag{\'o}n-Calvo} {et~al.}(2010{\natexlab{b}}){Arag{\'o}n-Calvo},
  {van de Weygaert}, \& {Jones}}]{Aragon-Calvo10}
{Arag{\'o}n-Calvo}, M.~A., {van de Weygaert}, R., \& {Jones}, B.~J.~T.
  2010{\natexlab{b}}, \mnras, 408, 2163

\bibitem[{{Arag{\'o}n-Calvo} {et~al.}(2007{\natexlab{b}}){Arag{\'o}n-Calvo},
  {van de Weygaert}, {Jones}, \& {van der Hulst}}]{Aragon-Calvo07b}
{Arag{\'o}n-Calvo}, M.~A., {van de Weygaert}, R., {Jones}, B.~J.~T., \& {van
  der Hulst}, J.~M. 2007{\natexlab{b}}, \apjl, 655, L5

\bibitem[{{Aragon-Calvo} {et~al.}(2015){Aragon-Calvo}, {Weygaert}, {Jones}, \&
  {Mobasher}}]{Aragon-calvo15}
{Aragon-Calvo}, M.~A., {Weygaert}, R.~v.~d., {Jones}, B.~J.~T., \& {Mobasher},
  B. 2015, \mnras, 454, 463

\bibitem[{{Balogh} {et~al.}(2000){Balogh}, {Navarro}, \& {Morris}}]{Balogh00}
{Balogh}, M.~L., {Navarro}, J.~F., \& {Morris}, S.~L. 2000, \apj, 540, 113

\bibitem[{{Becker} {et~al.}(2015){Becker}, {Bolton}, \& {Lidz}}]{Becker15}
{Becker}, G.~D., {Bolton}, J.~S., \& {Lidz}, A. 2015, \pasa, 32, e045

\bibitem[{{Berti} {et~al.}(2016){Berti}, {Coil}, {Behroozi}, {Eisenstein},
  {Bray}, {Cool}, \& {Moustakas}}]{Berti16}
{Berti}, A.~M., {Coil}, A.~L., {Behroozi}, P.~S., {et~al.} 2016, ArXiv
  e-prints, arXiv:1608.05084

\bibitem[{{Beygu} {et~al.}(2016){Beygu}, {Kreckel}, {van der Hulst}, {Jarrett},
  {Peletier}, {van de Weygaert}, {van Gorkom}, \& {Aragon-Calvo}}]{Beygu16}
{Beygu}, B., {Kreckel}, K., {van der Hulst}, J.~M., {et~al.} 2016, \mnras, 458,
  394

\bibitem[{{Biviano} {et~al.}(2011){Biviano}, {Fadda}, {Durret}, {Edwards}, \&
  {Marleau}}]{Biviano11}
{Biviano}, A., {Fadda}, D., {Durret}, F., {Edwards}, L.~O.~V., \& {Marleau}, F.
  2011, \aap, 532, A77

\bibitem[{{Bond} {et~al.}(1996){Bond}, {Kofman}, \& {Pogosyan}}]{Bond96}
{Bond}, J.~R., {Kofman}, L., \& {Pogosyan}, D. 1996, \nat, 380, 603

\bibitem[{{Bond} {et~al.}(2010){Bond}, {Strauss}, \& {Cen}}]{Bond10}
{Bond}, N.~A., {Strauss}, M.~A., \& {Cen}, R. 2010, \mnras, 409, 156

\bibitem[{{Brouwer} {et~al.}(2016){Brouwer}, {Cacciato}, {Dvornik}, {Eardley},
  {Heymans}, {Hoekstra}, {Kuijken}, {McNaught-Roberts}, {Sif{\'o}n}, {Viola},
  {Alpaslan}, {Bilicki}, {Bland-Hawthorn}, {Brough}, {Choi}, {Driver}, {Erben},
  {Grado}, {Hildebrandt}, {Holwerda}, {Hopkins}, {de Jong}, {Liske},
  {McFarland}, {Nakajima}, {Napolitano}, {Norberg}, {Peacock}, {Radovich},
  {Robotham}, {Schneider}, {Sikkema}, {van Uitert}, {Verdoes Kleijn}, \&
  {Valentijn}}]{Brouwer16}
{Brouwer}, M.~M., {Cacciato}, M., {Dvornik}, A., {et~al.} 2016, \mnras, 462,
  4451

\bibitem[{{Bruzual} \& {Charlot}(2003)}]{Bruzual03}
{Bruzual}, G., \& {Charlot}, S. 2003, \mnras, 344, 1000

\bibitem[{{Cantalupo} {et~al.}(2014){Cantalupo}, {Arrigoni-Battaia},
  {Prochaska}, {Hennawi}, \& {Madau}}]{Cantalupo14}
{Cantalupo}, S., {Arrigoni-Battaia}, F., {Prochaska}, J.~X., {Hennawi}, J.~F.,
  \& {Madau}, P. 2014, \nat, 506, 63

\bibitem[{{Capak} {et~al.}(2007){Capak}, {Aussel}, {Ajiki}, {McCracken},
  {Mobasher}, {Scoville}, {Shopbell}, {Taniguchi}, {Thompson}, {Tribiano},
  {Sasaki}, {Blain}, {Brusa}, {Carilli}, {Comastri}, {Carollo}, {Cassata},
  {Colbert}, {Ellis}, {Elvis}, {Giavalisco}, {Green}, {Guzzo}, {Hasinger},
  {Ilbert}, {Impey}, {Jahnke}, {Kartaltepe}, {Kneib}, {Koda}, {Koekemoer},
  {Komiyama}, {Leauthaud}, {Le Fevre}, {Lilly}, {Liu}, {Massey}, {Miyazaki},
  {Murayama}, {Nagao}, {Peacock}, {Pickles}, {Porciani}, {Renzini}, {Rhodes},
  {Rich}, {Salvato}, {Sanders}, {Scarlata}, {Schiminovich}, {Schinnerer},
  {Scodeggio}, {Sheth}, {Shioya}, {Tasca}, {Taylor}, {Yan}, \&
  {Zamorani}}]{Capak07}
{Capak}, P., {Aussel}, H., {Ajiki}, M., {et~al.} 2007, \apjs, 172, 99

\bibitem[{{Cautun} {et~al.}(2013){Cautun}, {van de Weygaert}, \&
  {Jones}}]{Cautun13}
{Cautun}, M., {van de Weygaert}, R., \& {Jones}, B.~J.~T. 2013, \mnras, 429,
  1286

\bibitem[{{Cautun} {et~al.}(2014){Cautun}, {van de Weygaert}, {Jones}, \&
  {Frenk}}]{Cautun14}
{Cautun}, M., {van de Weygaert}, R., {Jones}, B.~J.~T., \& {Frenk}, C.~S. 2014,
  \mnras, 441, 2923

\bibitem[{{Cen}(2014)}]{Cen14}
{Cen}, R. 2014, \apj, 781, 38

\bibitem[{{Cen} \& {Ostriker}(1999)}]{Cen99}
{Cen}, R., \& {Ostriker}, J.~P. 1999, \apj, 514, 1

\bibitem[{{Chabrier}(2003)}]{Chabrier03}
{Chabrier}, G. 2003, \pasp, 115, 763

\bibitem[{{Chen} {et~al.}(2016){Chen}, {Wang}, {Mo}, \& {Shi}}]{Chen16}
{Chen}, S., {Wang}, H., {Mo}, H.~J., \& {Shi}, J. 2016, \apj, 825, 49

\bibitem[{{Chen} {et~al.}(2015{\natexlab{a}}){Chen}, {Ho}, {Mandelbaum},
  {Bahcall}, {Brownstein}, {Freeman}, {Genovese}, {Schneider}, \&
  {Wasserman}}]{Chen15b}
{Chen}, Y.-C., {Ho}, S., {Mandelbaum}, R., {et~al.} 2015{\natexlab{a}}, ArXiv
  e-prints, arXiv:1509.06376

\bibitem[{{Chen} {et~al.}(2015{\natexlab{b}}){Chen}, {Ho}, {Tenneti},
  {Mandelbaum}, {Croft}, {DiMatteo}, {Freeman}, {Genovese}, \&
  {Wasserman}}]{Chen15}
{Chen}, Y.-C., {Ho}, S., {Tenneti}, A., {et~al.} 2015{\natexlab{b}}, \mnras,
  454, 3341

\bibitem[{{Codis} {et~al.}(2012){Codis}, {Pichon}, {Devriendt}, {Slyz},
  {Pogosyan}, {Dubois}, \& {Sousbie}}]{Codis12}
{Codis}, S., {Pichon}, C., {Devriendt}, J., {et~al.} 2012, \mnras, 427, 3320

\bibitem[{{Codis} {et~al.}(2015){Codis}, {Pichon}, \& {Pogosyan}}]{Codis15}
{Codis}, S., {Pichon}, C., \& {Pogosyan}, D. 2015, \mnras, 452, 3369

\bibitem[{{Colberg}(2007)}]{Colberg07}
{Colberg}, J.~M. 2007, \mnras, 375, 337

\bibitem[{{Colless} {et~al.}(2003){Colless}, {Peterson}, {Jackson}, {Peacock},
  {Cole}, {Norberg}, {Baldry}, {Baugh}, {Bland-Hawthorn}, {Bridges}, {Cannon},
  {Collins}, {Couch}, {Cross}, {Dalton}, {De Propris}, {Driver}, {Efstathiou},
  {Ellis}, {Frenk}, {Glazebrook}, {Lahav}, {Lewis}, {Lumsden}, {Maddox},
  {Madgwick}, {Sutherland}, \& {Taylor}}]{Colless03}
{Colless}, M., {Peterson}, B.~A., {Jackson}, C., {et~al.} 2003, ArXiv
  Astrophysics e-prints, astro-ph/0306581

\bibitem[{{Cooper} {et~al.}(2005){Cooper}, {Newman}, {Madgwick}, {Gerke},
  {Yan}, \& {Davis}}]{Cooper05}
{Cooper}, M.~C., {Newman}, J.~A., {Madgwick}, D.~S., {et~al.} 2005, \apj, 634,
  833

\bibitem[{{Darvish} {et~al.}(2015{\natexlab{a}}){Darvish}, {Mobasher},
  {Sobral}, {Hemmati}, {Nayyeri}, \& {Shivaei}}]{Darvish15b}
{Darvish}, B., {Mobasher}, B., {Sobral}, D., {et~al.} 2015{\natexlab{a}}, \apj,
  814, 84

\bibitem[{{Darvish} {et~al.}(2016){Darvish}, {Mobasher}, {Sobral}, {Rettura},
  {Scoville}, {Faisst}, \& {Capak}}]{Darvish16}
---. 2016, \apj, 825, 113

\bibitem[{{Darvish} {et~al.}(2015{\natexlab{b}}){Darvish}, {Mobasher},
  {Sobral}, {Scoville}, \& {Aragon-Calvo}}]{Darvish15a}
{Darvish}, B., {Mobasher}, B., {Sobral}, D., {Scoville}, N., \& {Aragon-Calvo},
  M. 2015{\natexlab{b}}, \apj, 805, 121

\bibitem[{{Darvish} {et~al.}(2014){Darvish}, {Sobral}, {Mobasher}, {Scoville},
  {Best}, {Sales}, \& {Smail}}]{Darvish14}
{Darvish}, B., {Sobral}, D., {Mobasher}, B., {et~al.} 2014, \apj, 796, 51

\bibitem[{{Dav{\'e}} {et~al.}(2001){Dav{\'e}}, {Cen}, {Ostriker}, {Bryan},
  {Hernquist}, {Katz}, {Weinberg}, {Norman}, \& {O'Shea}}]{Dave01}
{Dav{\'e}}, R., {Cen}, R., {Ostriker}, J.~P., {et~al.} 2001, \apj, 552, 473

\bibitem[{{Davis} {et~al.}(1985){Davis}, {Efstathiou}, {Frenk}, \&
  {White}}]{Davis85}
{Davis}, M., {Efstathiou}, G., {Frenk}, C.~S., \& {White}, S.~D.~M. 1985, \apj,
  292, 371

\bibitem[{{Dekel} {et~al.}(2009){Dekel}, {Birnboim}, {Engel}, {Freundlich},
  {Goerdt}, {Mumcuoglu}, {Neistein}, {Pichon}, {Teyssier}, \&
  {Zinger}}]{Dekel09a}
{Dekel}, A., {Birnboim}, Y., {Engel}, G., {et~al.} 2009, \nat, 457, 451

\bibitem[{{Doroshkevich} {et~al.}(2004){Doroshkevich}, {Tucker}, {Allam}, \&
  {Way}}]{Doroshkevich04}
{Doroshkevich}, A., {Tucker}, D.~L., {Allam}, S., \& {Way}, M.~J. 2004, \aap,
  418, 7

\bibitem[{{Duarte} \& {Mamon}(2014)}]{Duarte14}
{Duarte}, M., \& {Mamon}, G.~A. 2014, \mnras, 440, 1763

\bibitem[{{Dubois} {et~al.}(2014){Dubois}, {Pichon}, {Welker}, {Le Borgne},
  {Devriendt}, {Laigle}, {Codis}, {Pogosyan}, {Arnouts}, {Benabed}, {Bertin},
  {Blaizot}, {Bouchet}, {Cardoso}, {Colombi}, {de Lapparent}, {Desjacques},
  {Gavazzi}, {Kassin}, {Kimm}, {McCracken}, {Milliard}, {Peirani}, {Prunet},
  {Rouberol}, {Silk}, {Slyz}, {Sousbie}, {Teyssier}, {Tresse}, {Treyer},
  {Vibert}, \& {Volonteri}}]{Dubois14}
{Dubois}, Y., {Pichon}, C., {Welker}, C., {et~al.} 2014, \mnras, 444, 1453

\bibitem[{{Eardley} {et~al.}(2015){Eardley}, {Peacock}, {McNaught-Roberts},
  {Heymans}, {Norberg}, {Alpaslan}, {Baldry}, {Bland-Hawthorn}, {Brough},
  {Cluver}, {Driver}, {Farrow}, {Liske}, {Loveday}, \& {Robotham}}]{Eardley15}
{Eardley}, E., {Peacock}, J.~A., {McNaught-Roberts}, T., {et~al.} 2015, \mnras,
  448, 3665

\bibitem[{{Erfanianfar} {et~al.}(2016){Erfanianfar}, {Popesso}, {Finoguenov},
  {Wilman}, {Wuyts}, {Biviano}, {Salvato}, {Mirkazemi}, {Morselli}, {Ziparo},
  {Nandra}, {Lutz}, {Elbaz}, {Dickinson}, {Tanaka}, {Altieri}, {Aussel},
  {Bauer}, {Berta}, {Bielby}, {Brandt}, {Cappelluti}, {Cimatti}, {Cooper},
  {Fadda}, {Ilbert}, {Le Floch}, {Magnelli}, {Mulchaey}, {Nordon}, {Newman},
  {Poglitsch}, \& {Pozzi}}]{Erfanianfar16}
{Erfanianfar}, G., {Popesso}, P., {Finoguenov}, A., {et~al.} 2016, \mnras, 455,
  2839

\bibitem[{{Fabian}(2012)}]{Fabian12}
{Fabian}, A.~C. 2012, \araa, 50, 455

\bibitem[{{Fadda} {et~al.}(2008){Fadda}, {Biviano}, {Marleau},
  {Storrie-Lombardi}, \& {Durret}}]{Fadda08}
{Fadda}, D., {Biviano}, A., {Marleau}, F.~R., {Storrie-Lombardi}, L.~J., \&
  {Durret}, F. 2008, \apjl, 672, L9

\bibitem[{{Falck} {et~al.}(2012){Falck}, {Neyrinck}, \& {Szalay}}]{Falck12}
{Falck}, B.~L., {Neyrinck}, M.~C., \& {Szalay}, A.~S. 2012, \apj, 754, 126

\bibitem[{{Farouki} \& {Shapiro}(1981)}]{Farouki81}
{Farouki}, R., \& {Shapiro}, S.~L. 1981, \apj, 243, 32

\bibitem[{{Filho} {et~al.}(2015){Filho}, {S{\'a}nchez Almeida},
  {Mu{\~n}oz-Tu{\~n}{\'o}n}, {Nuza}, {Kitaura}, \& {He{\ss}}}]{Filho15}
{Filho}, M.~E., {S{\'a}nchez Almeida}, J., {Mu{\~n}oz-Tu{\~n}{\'o}n}, C.,
  {et~al.} 2015, \apj, 802, 82

\bibitem[{{Finoguenov} {et~al.}(2007){Finoguenov}, {Guzzo}, {Hasinger},
  {Scoville}, {Aussel}, {B{\"o}hringer}, {Brusa}, {Capak}, {Cappelluti},
  {Comastri}, {Giodini}, {Griffiths}, {Impey}, {Koekemoer}, {Kneib},
  {Leauthaud}, {Le F{\`e}vre}, {Lilly}, {Mainieri}, {Massey}, {McCracken},
  {Mobasher}, {Murayama}, {Peacock}, {Sakelliou}, {Schinnerer}, {Silverman},
  {Smol{\v c}i{\'c}}, {Taniguchi}, {Tasca}, {Taylor}, {Trump}, \&
  {Zamorani}}]{Finoguenov07}
{Finoguenov}, A., {Guzzo}, L., {Hasinger}, G., {et~al.} 2007, \apjs, 172, 182

\bibitem[{{Forero-Romero} {et~al.}(2009){Forero-Romero}, {Hoffman},
  {Gottl{\"o}ber}, {Klypin}, \& {Yepes}}]{Forero-romero09}
{Forero-Romero}, J.~E., {Hoffman}, Y., {Gottl{\"o}ber}, S., {Klypin}, A., \&
  {Yepes}, G. 2009, \mnras, 396, 1815

\bibitem[{{Frangi} {et~al.}(1998){Frangi}, {Niessen}, {Vincken}, \&
  {Viergever}}]{Frangi98}
{Frangi}, A.~F., {Niessen}, W.~J., {Vincken}, K.~L., \& {Viergever}, M.~A.
  1998, {}

\bibitem[{{Geller} \& {Huchra}(1989)}]{Geller89}
{Geller}, M.~J., \& {Huchra}, J.~P. 1989, Science, 246, 897

\bibitem[{{George} {et~al.}(2011){George}, {Leauthaud}, {Bundy}, {Finoguenov},
  {Tinker}, {Lin}, {Mei}, {Kneib}, {Aussel}, {Behroozi}, {Busha}, {Capak},
  {Coccato}, {Covone}, {Faure}, {Fiorenza}, {Ilbert}, {Le Floc'h}, {Koekemoer},
  {Tanaka}, {Wechsler}, \& {Wolk}}]{George11}
{George}, M.~R., {Leauthaud}, A., {Bundy}, K., {et~al.} 2011, \apj, 742, 125

\bibitem[{{Gonz{\'a}lez} \& {Padilla}(2010)}]{Gonzalez10}
{Gonz{\'a}lez}, R.~E., \& {Padilla}, N.~D. 2010, \mnras, 407, 1449

\bibitem[{{Gonzalez} {et~al.}(2016){Gonzalez}, {Prieto}, {Padilla}, \&
  {Jimenez}}]{Gonzalez16}
{Gonzalez}, R.~E., {Prieto}, J., {Padilla}, N., \& {Jimenez}, R. 2016, ArXiv
  e-prints, arXiv:1606.04562

\bibitem[{{Gunn} \& {Gott}(1972)}]{Gunn72}
{Gunn}, J.~E., \& {Gott}, III, J.~R. 1972, \apj, 176, 1

\bibitem[{{Guo} {et~al.}(2015){Guo}, {Tempel}, \& {Libeskind}}]{Guo15}
{Guo}, Q., {Tempel}, E., \& {Libeskind}, N.~I. 2015, \apj, 800, 112

\bibitem[{{Hahn} {et~al.}(2016){Hahn}, {Tinker}, \& {Wetzel}}]{Hahn16}
{Hahn}, C., {Tinker}, J.~L., \& {Wetzel}, A.~R. 2016, ArXiv e-prints,
  arXiv:1609.04398

\bibitem[{{Hahn} {et~al.}(2007{\natexlab{a}}){Hahn}, {Carollo}, {Porciani}, \&
  {Dekel}}]{Hahn07b}
{Hahn}, O., {Carollo}, C.~M., {Porciani}, C., \& {Dekel}, A.
  2007{\natexlab{a}}, \mnras, 381, 41

\bibitem[{{Hahn} {et~al.}(2007{\natexlab{b}}){Hahn}, {Porciani}, {Carollo}, \&
  {Dekel}}]{Hahn07}
{Hahn}, O., {Porciani}, C., {Carollo}, C.~M., \& {Dekel}, A.
  2007{\natexlab{b}}, \mnras, 375, 489

\bibitem[{{Haider} {et~al.}(2016){Haider}, {Steinhauser}, {Vogelsberger},
  {Genel}, {Springel}, {Torrey}, \& {Hernquist}}]{Haider16}
{Haider}, M., {Steinhauser}, D., {Vogelsberger}, M., {et~al.} 2016, \mnras,
  457, 3024

\bibitem[{{Haines} {et~al.}(2013){Haines}, {Pereira}, {Smith}, {Egami},
  {Sanderson}, {Babul}, {Finoguenov}, {Merluzzi}, {Busarello}, {Rawle}, \&
  {Okabe}}]{Haines13}
{Haines}, C.~P., {Pereira}, M.~J., {Smith}, G.~P., {et~al.} 2013, \apj, 775,
  126

\bibitem[{{Hartley} {et~al.}(2015){Hartley}, {Conselice}, {Mortlock},
  {Foucaud}, \& {Simpson}}]{Hartley15}
{Hartley}, W.~G., {Conselice}, C.~J., {Mortlock}, A., {Foucaud}, S., \&
  {Simpson}, C. 2015, \mnras, 451, 1613

\bibitem[{{Hatfield} \& {Jarvis}(2016)}]{Hatfield16}
{Hatfield}, P.~W., \& {Jarvis}, M.~J. 2016, ArXiv e-prints, arXiv:1606.08989

\bibitem[{{Hearin} {et~al.}(2015){Hearin}, {Watson}, \& {van den
  Bosch}}]{Hearin15}
{Hearin}, A.~P., {Watson}, D.~F., \& {van den Bosch}, F.~C. 2015, \mnras, 452,
  1958

\bibitem[{{Hoffman} {et~al.}(2012){Hoffman}, {Metuki}, {Yepes},
  {Gottl{\"o}ber}, {Forero-Romero}, {Libeskind}, \& {Knebe}}]{Hoffman12}
{Hoffman}, Y., {Metuki}, O., {Yepes}, G., {et~al.} 2012, \mnras, 425, 2049

\bibitem[{{Hopkins} {et~al.}(2014){Hopkins}, {Kere{\v s}}, {O{\~n}orbe},
  {Faucher-Gigu{\`e}re}, {Quataert}, {Murray}, \& {Bullock}}]{Hopkins14}
{Hopkins}, P.~F., {Kere{\v s}}, D., {O{\~n}orbe}, J., {et~al.} 2014, \mnras,
  445, 581

\bibitem[{{Huchra} \& {Geller}(1982)}]{Huchra82}
{Huchra}, J.~P., \& {Geller}, M.~J. 1982, \apj, 257, 423

\bibitem[{{Ilbert} {et~al.}(2009){Ilbert}, {Capak}, {Salvato}, {Aussel},
  {McCracken}, {Sanders}, {Scoville}, {Kartaltepe}, {Arnouts}, {Le Floc'h},
  {Mobasher}, {Taniguchi}, {Lamareille}, {Leauthaud}, {Sasaki}, {Thompson},
  {Zamojski}, {Zamorani}, {Bardelli}, {Bolzonella}, {Bongiorno}, {Brusa},
  {Caputi}, {Carollo}, {Contini}, {Cook}, {Coppa}, {Cucciati}, {de la Torre},
  {de Ravel}, {Franzetti}, {Garilli}, {Hasinger}, {Iovino}, {Kampczyk},
  {Kneib}, {Knobel}, {Kovac}, {Le Borgne}, {Le Brun}, {F{\`e}vre}, {Lilly},
  {Looper}, {Maier}, {Mainieri}, {Mellier}, {Mignoli}, {Murayama}, {Pell{\`o}},
  {Peng}, {P{\'e}rez-Montero}, {Renzini}, {Ricciardelli}, {Schiminovich},
  {Scodeggio}, {Shioya}, {Silverman}, {Surace}, {Tanaka}, {Tasca}, {Tresse},
  {Vergani}, \& {Zucca}}]{Ilbert09}
{Ilbert}, O., {Capak}, P., {Salvato}, M., {et~al.} 2009, \apj, 690, 1236

\bibitem[{{Ilbert} {et~al.}(2013){Ilbert}, {McCracken}, {Le F{\`e}vre},
  {Capak}, {Dunlop}, {Karim}, {Renzini}, {Caputi}, {Boissier}, {Arnouts},
  {Aussel}, {Comparat}, {Guo}, {Hudelot}, {Kartaltepe}, {Kneib}, {Krogager},
  {Le Floc'h}, {Lilly}, {Mellier}, {Milvang-Jensen}, {Moutard}, {Onodera},
  {Richard}, {Salvato}, {Sanders}, {Scoville}, {Silverman}, {Taniguchi},
  {Tasca}, {Thomas}, {Toft}, {Tresse}, {Vergani}, {Wolk}, \& {Zirm}}]{Ilbert13}
{Ilbert}, O., {McCracken}, H.~J., {Le F{\`e}vre}, O., {et~al.} 2013, \aap, 556,
  A55

\bibitem[{{Ilbert} {et~al.}(2015){Ilbert}, {Arnouts}, {Le Floc'h}, {Aussel},
  {Bethermin}, {Capak}, {Hsieh}, {Kajisawa}, {Karim}, {Le F{\`e}vre}, {Lee},
  {Lilly}, {McCracken}, {Michel-Dansac}, {Moutard}, {Renzini}, {Salvato},
  {Sanders}, {Scoville}, {Sheth}, {Silverman}, {Smol{\v c}i{\'c}}, {Taniguchi},
  \& {Tresse}}]{Ilbert15}
{Ilbert}, O., {Arnouts}, S., {Le Floc'h}, E., {et~al.} 2015, \aap, 579, A2

\bibitem[{{Jarrett}(2004)}]{Jarrett04}
{Jarrett}, T. 2004, \pasa, 21, 396

\bibitem[{{Jasche} {et~al.}(2010){Jasche}, {Kitaura}, {Li}, \&
  {En{\ss}lin}}]{Jasche10}
{Jasche}, J., {Kitaura}, F.~S., {Li}, C., \& {En{\ss}lin}, T.~A. 2010, \mnras,
  409, 355

\bibitem[{{Joachimi} {et~al.}(2015){Joachimi}, {Cacciato}, {Kitching},
  {Leonard}, {Mandelbaum}, {Sch{\"a}fer}, {Sif{\'o}n}, {Hoekstra}, {Kiessling},
  {Kirk}, \& {Rassat}}]{Joachimi15}
{Joachimi}, B., {Cacciato}, M., {Kitching}, T.~D., {et~al.} 2015, \ssr, 193, 1

\bibitem[{{Jones} {et~al.}(2010){Jones}, {van de Weygaert}, \&
  {Arag{\'o}n-Calvo}}]{Jones10}
{Jones}, B.~J.~T., {van de Weygaert}, R., \& {Arag{\'o}n-Calvo}, M.~A. 2010,
  \mnras, 408, 897

\bibitem[{{Jones} {et~al.}(2009){Jones}, {Read}, {Saunders}, {Colless},
  {Jarrett}, {Parker}, {Fairall}, {Mauch}, {Sadler}, {Watson}, {Burton},
  {Campbell}, {Cass}, {Croom}, {Dawe}, {Fiegert}, {Frankcombe}, {Hartley},
  {Huchra}, {James}, {Kirby}, {Lahav}, {Lucey}, {Mamon}, {Moore}, {Peterson},
  {Prior}, {Proust}, {Russell}, {Safouris}, {Wakamatsu}, {Westra}, \&
  {Williams}}]{Jones09}
{Jones}, D.~H., {Read}, M.~A., {Saunders}, W., {et~al.} 2009, \mnras, 399, 683

\bibitem[{{Kang} \& {Wang}(2015)}]{Kang15}
{Kang}, X., \& {Wang}, P. 2015, \apj, 813, 6

\bibitem[{{Kashikawa} \& {Okamura}(1992)}]{Kashikawa92}
{Kashikawa}, N., \& {Okamura}, S. 1992, \pasj, 44, 493

\bibitem[{{Kauffmann} {et~al.}(2013){Kauffmann}, {Li}, {Zhang}, \&
  {Weinmann}}]{Kauffmann13}
{Kauffmann}, G., {Li}, C., {Zhang}, W., \& {Weinmann}, S. 2013, \mnras, 430,
  1447

\bibitem[{{Kawinwanichakij} {et~al.}(2016){Kawinwanichakij}, {Quadri},
  {Papovich}, {Kacprzak}, {Labb{\'e}}, {Spitler}, {Straatman}, {Tran}, {Allen},
  {Behroozi}, {Cowley}, {Dekel}, {Glazebrook}, {Hartley}, {Kelson}, {Koo},
  {Lee}, {Lu}, {Nanayakkara}, {Persson}, {Primack}, {Tilvi}, {Tomczak}, \& {van
  Dokkum}}]{Kawinwanichakij16}
{Kawinwanichakij}, L., {Quadri}, R.~F., {Papovich}, C., {et~al.} 2016, \apj,
  817, 9

\bibitem[{{Kere{\v s}} {et~al.}(2005){Kere{\v s}}, {Katz}, {Weinberg}, \&
  {Dav{\'e}}}]{Keres05}
{Kere{\v s}}, D., {Katz}, N., {Weinberg}, D.~H., \& {Dav{\'e}}, R. 2005,
  \mnras, 363, 2

\bibitem[{{Kiessling} {et~al.}(2015){Kiessling}, {Cacciato}, {Joachimi},
  {Kirk}, {Kitching}, {Leonard}, {Mandelbaum}, {Sch{\"a}fer}, {Sif{\'o}n},
  {Brown}, \& {Rassat}}]{Kiessling15}
{Kiessling}, A., {Cacciato}, M., {Joachimi}, B., {et~al.} 2015, \ssr, 193, 67

\bibitem[{{Kova{\v c}} {et~al.}(2014){Kova{\v c}}, {Lilly}, {Knobel},
  {Bschorr}, {Peng}, {Carollo}, {Contini}, {Kneib}, {Le F{\'e}vre}, {Mainieri},
  {Renzini}, {Scodeggio}, {Zamorani}, {Bardelli}, {Bolzonella}, {Bongiorno},
  {Caputi}, {Cucciati}, {de la Torre}, {de Ravel}, {Franzetti}, {Garilli},
  {Iovino}, {Kampczyk}, {Lamareille}, {Le Borgne}, {Le Brun}, {Maier},
  {Mignoli}, {Oesch}, {Pello}, {Montero}, {Presotto}, {Silverman}, {Tanaka},
  {Tasca}, {Tresse}, {Vergani}, {Zucca}, {Aussel}, {Koekemoer}, {Le Floc'h},
  {Moresco}, \& {Pozzetti}}]{Kovac14}
{Kova{\v c}}, K., {Lilly}, S.~J., {Knobel}, C., {et~al.} 2014, \mnras, 438, 717

\bibitem[{{Koyama} {et~al.}(2013){Koyama}, {Smail}, {Kurk}, {Geach}, {Sobral},
  {Kodama}, {Nakata}, {Swinbank}, {Best}, {Hayashi}, \& {Tadaki}}]{Koyama13a}
{Koyama}, Y., {Smail}, I., {Kurk}, J., {et~al.} 2013, \mnras, 434, 423

\bibitem[{{Lai} {et~al.}(2016){Lai}, {Lin}, {Jian}, {Chiueh}, {Merson},
  {Baugh}, {Foucaud}, {Chen}, \& {Chen}}]{Lai16}
{Lai}, C.-C., {Lin}, L., {Jian}, H.-Y., {et~al.} 2016, \apj, 825, 40

\bibitem[{{Laigle} {et~al.}(2016){Laigle}, {McCracken}, {Ilbert}, {Hsieh},
  {Davidzon}, {Capak}, {Hasinger}, {Silverman}, {Pichon}, {Coupon}, {Aussel},
  {Le Borgne}, {Caputi}, {Cassata}, {Chang}, {Civano}, {Dunlop}, {Fynbo},
  {kartaltepe}, {Koekemoer}, {Le Fevre}, {Le Floc'h}, {Leauthaud}, {Lilly},
  {Lin}, {Marchesi}, {Milvang-Jensen}, {Salvato}, {Sanders}, {Scoville},
  {Smolcic}, {Stockmann}, {Taniguchi}, {Tasca}, {Toft}, {Vaccari}, \&
  {Zabl}}]{Laigle16}
{Laigle}, C., {McCracken}, H.~J., {Ilbert}, O., {et~al.} 2016, ArXiv e-prints,
  arXiv:1604.02350

\bibitem[{{Larson} {et~al.}(1980){Larson}, {Tinsley}, \& {Caldwell}}]{Larson80}
{Larson}, R.~B., {Tinsley}, B.~M., \& {Caldwell}, C.~N. 1980, \apj, 237, 692

\bibitem[{{Leclercq} {et~al.}(2015){Leclercq}, {Jasche}, \&
  {Wandelt}}]{Leclercq15}
{Leclercq}, F., {Jasche}, J., \& {Wandelt}, B. 2015, \jcap, 6, 015

\bibitem[{{Lee} \& {Erdogdu}(2007)}]{Lee07}
{Lee}, J., \& {Erdogdu}, P. 2007, \apj, 671, 1248

\bibitem[{{Lee} {et~al.}(2013){Lee}, {Sanders}, {Casey}, {Scoville}, {Hung},
  {Le Floc'h}, {Ilbert}, {Aussel}, {Capak}, {Kartaltepe}, {Roseboom},
  {Salvato}, {Aravena}, {Berta}, {Bock}, {Oliver}, {Riguccini}, \&
  {Symeonidis}}]{Lee13}
{Lee}, N., {Sanders}, D.~B., {Casey}, C.~M., {et~al.} 2013, \apj, 778, 131

\bibitem[{{Lee} {et~al.}(2015){Lee}, {Sanders}, {Casey}, {Toft}, {Scoville},
  {Hung}, {Le Floc'h}, {Ilbert}, {Zahid}, {Aussel}, {Capak}, {Kartaltepe},
  {Kewley}, {Li}, {Schawinski}, {Sheth}, \& {Xiao}}]{Lee15a}
---. 2015, \apj, 801, 80

\bibitem[{{Libeskind} {et~al.}(2013){Libeskind}, {Hoffman}, {Forero-Romero},
  {Gottl{\"o}ber}, {Knebe}, {Steinmetz}, \& {Klypin}}]{Libeskind13}
{Libeskind}, N.~I., {Hoffman}, Y., {Forero-Romero}, J., {et~al.} 2013, \mnras,
  428, 2489

\bibitem[{{Lilly} {et~al.}(2009){Lilly}, {Le Brun}, {Maier}, {Mainieri},
  {Mignoli}, {Scodeggio}, {Zamorani}, {Carollo}, {Contini}, {Kneib}, {Le
  F{\`e}vre}, {Renzini}, {Bardelli}, {Bolzonella}, {Bongiorno}, {Caputi},
  {Coppa}, {Cucciati}, {de la Torre}, {de Ravel}, {Franzetti}, {Garilli},
  {Iovino}, {Kampczyk}, {Kovac}, {Knobel}, {Lamareille}, {Le Borgne}, {Pello},
  {Peng}, {P{\'e}rez-Montero}, {Ricciardelli}, {Silverman}, {Tanaka}, {Tasca},
  {Tresse}, {Vergani}, {Zucca}, {Ilbert}, {Salvato}, {Oesch}, {Abbas},
  {Bottini}, {Capak}, {Cappi}, {Cassata}, {Cimatti}, {Elvis}, {Fumana},
  {Guzzo}, {Hasinger}, {Koekemoer}, {Leauthaud}, {Maccagni}, {Marinoni},
  {McCracken}, {Memeo}, {Meneux}, {Porciani}, {Pozzetti}, {Sanders},
  {Scaramella}, {Scarlata}, {Scoville}, {Shopbell}, \& {Taniguchi}}]{Lilly09}
{Lilly}, S.~J., {Le Brun}, V., {Maier}, C., {et~al.} 2009, \apjs, 184, 218

\bibitem[{{Lin} {et~al.}(2014){Lin}, {Jian}, {Foucaud}, {Norberg}, {Bower},
  {Cole}, {Arnalte-Mur}, {Chen}, {Coupon}, {Hsieh}, {Heinis}, {Phleps}, {Chen},
  {Lee}, {Burgett}, {Chambers}, {Denneau}, {Draper}, {Flewelling}, {Hodapp},
  {Huber}, {Kaiser}, {Kudritzki}, {Magnier}, {Metcalfe}, {Price}, {Tonry},
  {Wainscoat}, \& {Waters}}]{Lin14}
{Lin}, L., {Jian}, H.-Y., {Foucaud}, S., {et~al.} 2014, \apj, 782, 33

\bibitem[{{Malavasi} {et~al.}(2016){Malavasi}, {Pozzetti}, {Cucciati},
  {Bardelli}, \& {Cimatti}}]{Malavasi16}
{Malavasi}, N., {Pozzetti}, L., {Cucciati}, O., {Bardelli}, S., \& {Cimatti},
  A. 2016, \aap, 585, A116

\bibitem[{{Martin} {et~al.}(2014{\natexlab{a}}){Martin}, {Chang},
  {Matuszewski}, {Morrissey}, {Rahman}, {Moore}, \& {Steidel}}]{Martin14b}
{Martin}, D.~C., {Chang}, D., {Matuszewski}, M., {et~al.} 2014{\natexlab{a}},
  \apj, 786, 106

\bibitem[{{Martin} {et~al.}(2014{\natexlab{b}}){Martin}, {Chang},
  {Matuszewski}, {Morrissey}, {Rahman}, {Moore}, {Steidel}, \&
  {Matsuda}}]{Martin14a}
---. 2014{\natexlab{b}}, \apj, 786, 107

\bibitem[{{Martin} {et~al.}(2015){Martin}, {Matuszewski}, {Morrissey}, {Neill},
  {Moore}, {Cantalupo}, {Prochaska}, \& {Chang}}]{Martin15}
{Martin}, D.~C., {Matuszewski}, M., {Morrissey}, P., {et~al.} 2015, \nat, 524,
  192

\bibitem[{{Martin} {et~al.}(2016){Martin}, {Matuszewski}, {Morrissey}, {Neill},
  {Moore}, {Steidel}, \& {Trainor}}]{Martin16}
---. 2016, \apjl, 824, L5

\bibitem[{{McCracken} {et~al.}(2012){McCracken}, {Milvang-Jensen}, {Dunlop},
  {Franx}, {Fynbo}, {Le F{\`e}vre}, {Holt}, {Caputi}, {Goranova}, {Buitrago},
  {Emerson}, {Freudling}, {Hudelot}, {L{\'o}pez-Sanjuan}, {Magnard}, {Mellier},
  {M{\o}ller}, {Nilsson}, {Sutherland}, {Tasca}, \& {Zabl}}]{McCracken12}
{McCracken}, H.~J., {Milvang-Jensen}, B., {Dunlop}, J., {et~al.} 2012, \aap,
  544, A156

\bibitem[{{Merritt}(1983)}]{Merritt83}
{Merritt}, D. 1983, \apj, 264, 24

\bibitem[{{Moore} {et~al.}(1998){Moore}, {Lake}, \& {Katz}}]{Moore98}
{Moore}, B., {Lake}, G., \& {Katz}, N. 1998, \apj, 495, 139

\bibitem[{{Moster} {et~al.}(2011){Moster}, {Somerville}, {Newman}, \&
  {Rix}}]{Moster11}
{Moster}, B.~P., {Somerville}, R.~S., {Newman}, J.~A., \& {Rix}, H.-W. 2011,
  \apj, 731, 113

\bibitem[{{Muzzin} {et~al.}(2012){Muzzin}, {Wilson}, {Yee}, {Gilbank},
  {Hoekstra}, {Demarco}, {Balogh}, {van Dokkum}, {Franx}, {Ellingson}, {Hicks},
  {Nantais}, {Noble}, {Lacy}, {Lidman}, {Rettura}, {Surace}, \&
  {Webb}}]{Muzzin12}
{Muzzin}, A., {Wilson}, G., {Yee}, H.~K.~C., {et~al.} 2012, \apj, 746, 188

\bibitem[{{Navarro} {et~al.}(2004){Navarro}, {Abadi}, \&
  {Steinmetz}}]{Navarro04}
{Navarro}, J.~F., {Abadi}, M.~G., \& {Steinmetz}, M. 2004, \apjl, 613, L41

\bibitem[{{Novikov} {et~al.}(2006){Novikov}, {Colombi}, \&
  {Dor{\'e}}}]{Novikov06}
{Novikov}, D., {Colombi}, S., \& {Dor{\'e}}, O. 2006, \mnras, 366, 1201

\bibitem[{{Pandey} \& {Sarkar}(2016)}]{Pandey16}
{Pandey}, B., \& {Sarkar}, S. 2016, ArXiv e-prints, arXiv:1611.00283

\bibitem[{{Patel} {et~al.}(2009){Patel}, {Holden}, {Kelson}, {Illingworth}, \&
  {Franx}}]{Patel09}
{Patel}, S.~G., {Holden}, B.~P., {Kelson}, D.~D., {Illingworth}, G.~D., \&
  {Franx}, M. 2009, \apjl, 705, L67

\bibitem[{{Patel} {et~al.}(2011){Patel}, {Kelson}, {Holden}, {Franx}, \&
  {Illingworth}}]{Patel11}
{Patel}, S.~G., {Kelson}, D.~D., {Holden}, B.~P., {Franx}, M., \&
  {Illingworth}, G.~D. 2011, \apj, 735, 53

\bibitem[{{Paz} {et~al.}(2008){Paz}, {Stasyszyn}, \& {Padilla}}]{Paz08}
{Paz}, D.~J., {Stasyszyn}, F., \& {Padilla}, N.~D. 2008, \mnras, 389, 1127

\bibitem[{{Peebles}(1969)}]{Peebles69}
{Peebles}, P.~J.~E. 1969, \apj, 155, 393

\bibitem[{{Peng} {et~al.}(2012){Peng}, {Lilly}, {Renzini}, \&
  {Carollo}}]{Peng12}
{Peng}, Y.-j., {Lilly}, S.~J., {Renzini}, A., \& {Carollo}, M. 2012, \apj, 757,
  4

\bibitem[{{Peng} {et~al.}(2010){Peng}, {Lilly}, {Kova{\v c}}, {Bolzonella},
  {Pozzetti}, {Renzini}, {Zamorani}, {Ilbert}, {Knobel}, {Iovino}, {Maier},
  {Cucciati}, {Tasca}, {Carollo}, {Silverman}, {Kampczyk}, {de Ravel},
  {Sanders}, {Scoville}, {Contini}, {Mainieri}, {Scodeggio}, {Kneib}, {Le
  F{\`e}vre}, {Bardelli}, {Bongiorno}, {Caputi}, {Coppa}, {de la Torre},
  {Franzetti}, {Garilli}, {Lamareille}, {Le Borgne}, {Le Brun}, {Mignoli},
  {Perez Montero}, {Pello}, {Ricciardelli}, {Tanaka}, {Tresse}, {Vergani},
  {Welikala}, {Zucca}, {Oesch}, {Abbas}, {Barnes}, {Bordoloi}, {Bottini},
  {Cappi}, {Cassata}, {Cimatti}, {Fumana}, {Hasinger}, {Koekemoer},
  {Leauthaud}, {Maccagni}, {Marinoni}, {McCracken}, {Memeo}, {Meneux}, {Nair},
  {Porciani}, {Presotto}, \& {Scaramella}}]{Peng10}
{Peng}, Y.-j., {Lilly}, S.~J., {Kova{\v c}}, K., {et~al.} 2010, \apj, 721, 193

\bibitem[{{Penny} {et~al.}(2015){Penny}, {Brown}, {Pimbblet}, {Cluver},
  {Croton}, {Owers}, {Lange}, {Alpaslan}, {Baldry}, {Bland-Hawthorn}, {Brough},
  {Driver}, {Holwerda}, {Hopkins}, {Jarrett}, {Jones}, {Kelvin},
  {Lara-L{\'o}pez}, {Liske}, {L{\'o}pez-S{\'a}nchez}, {Loveday}, {Meyer},
  {Norberg}, {Robotham}, \& {Rodrigues}}]{Penny15}
{Penny}, S.~J., {Brown}, M.~J.~I., {Pimbblet}, K.~A., {et~al.} 2015, \mnras,
  453, 3519

\bibitem[{{Pimbblet}(2005)}]{Pimbblet05}
{Pimbblet}, K.~A. 2005, \mnras, 358, 256

\bibitem[{{Porter} {et~al.}(2008){Porter}, {Raychaudhury}, {Pimbblet}, \&
  {Drinkwater}}]{Porter08}
{Porter}, S.~C., {Raychaudhury}, S., {Pimbblet}, K.~A., \& {Drinkwater}, M.~J.
  2008, \mnras, 388, 1152

\bibitem[{{Poudel} {et~al.}(2016){Poudel}, {Hein{\"a}m{\"a}ki}, {Tempel},
  {Einasto}, {Lietzen}, \& {Nurmi}}]{Poudel16}
{Poudel}, A., {Hein{\"a}m{\"a}ki}, P., {Tempel}, E., {et~al.} 2016, ArXiv
  e-prints, arXiv:1611.01072

\bibitem[{{Pozzetti} {et~al.}(2010){Pozzetti}, {Bolzonella}, {Zucca},
  {Zamorani}, {Lilly}, {Renzini}, {Moresco}, {Mignoli}, {Cassata}, {Tasca},
  {Lamareille}, {Maier}, {Meneux}, {Halliday}, {Oesch}, {Vergani}, {Caputi},
  {Kova{\v c}}, {Cimatti}, {Cucciati}, {Iovino}, {Peng}, {Carollo}, {Contini},
  {Kneib}, {Le F{\'e}vre}, {Mainieri}, {Scodeggio}, {Bardelli}, {Bongiorno},
  {Coppa}, {de la Torre}, {de Ravel}, {Franzetti}, {Garilli}, {Kampczyk},
  {Knobel}, {Le Borgne}, {Le Brun}, {Pell{\`o}}, {Perez Montero},
  {Ricciardelli}, {Silverman}, {Tanaka}, {Tresse}, {Abbas}, {Bottini}, {Cappi},
  {Guzzo}, {Koekemoer}, {Leauthaud}, {Maccagni}, {Marinoni}, {McCracken},
  {Memeo}, {Porciani}, {Scaramella}, {Scarlata}, \& {Scoville}}]{Pozzetti10}
{Pozzetti}, L., {Bolzonella}, M., {Zucca}, E., {et~al.} 2010, \aap, 523, A13

\bibitem[{{Quadri} {et~al.}(2012){Quadri}, {Williams}, {Franx}, \&
  {Hildebrandt}}]{Quadri12}
{Quadri}, R.~F., {Williams}, R.~J., {Franx}, M., \& {Hildebrandt}, H. 2012,
  \apj, 744, 88

\bibitem[{{Ricciardelli} {et~al.}(2014){Ricciardelli}, {Cava}, {Varela}, \&
  {Quilis}}]{Ricciardelli14}
{Ricciardelli}, E., {Cava}, A., {Varela}, J., \& {Quilis}, V. 2014, \mnras,
  445, 4045

\bibitem[{{Scoville} {et~al.}(2007{\natexlab{a}}){Scoville}, {Aussel},
  {Benson}, {Blain}, {Calzetti}, {Capak}, {Ellis}, {El-Zant}, {Finoguenov},
  {Giavalisco}, {Guzzo}, {Hasinger}, {Koda}, {Le F{\`e}vre}, {Massey},
  {McCracken}, {Mobasher}, {Renzini}, {Rhodes}, {Salvato}, {Sanders}, {Sasaki},
  {Schinnerer}, {Sheth}, {Shopbell}, {Taniguchi}, {Taylor}, \&
  {Thompson}}]{Scoville07b}
{Scoville}, N., {Aussel}, H., {Benson}, A., {et~al.} 2007{\natexlab{a}}, \apjs,
  172, 150

\bibitem[{{Scoville} {et~al.}(2007{\natexlab{b}}){Scoville}, {Aussel}, {Brusa},
  {Capak}, {Carollo}, {Elvis}, {Giavalisco}, {Guzzo}, {Hasinger}, {Impey},
  {Kneib}, {LeFevre}, {Lilly}, {Mobasher}, {Renzini}, {Rich}, {Sanders},
  {Schinnerer}, {Schminovich}, {Shopbell}, {Taniguchi}, \&
  {Tyson}}]{Scoville07}
{Scoville}, N., {Aussel}, H., {Brusa}, M., {et~al.} 2007{\natexlab{b}}, \apjs,
  172, 1

\bibitem[{{Scoville} {et~al.}(2013){Scoville}, {Arnouts}, {Aussel}, {Benson},
  {Bongiorno}, {Bundy}, {Calvo}, {Capak}, {Carollo}, {Civano}, {Dunlop},
  {Elvis}, {Faisst}, {Finoguenov}, {Fu}, {Giavalisco}, {Guo}, {Ilbert},
  {Iovino}, {Kajisawa}, {Kartaltepe}, {Leauthaud}, {Le F{\`e}vre}, {LeFloch},
  {Lilly}, {Liu}, {Manohar}, {Massey}, {Masters}, {McCracken}, {Mobasher},
  {Peng}, {Renzini}, {Rhodes}, {Salvato}, {Sanders}, {Sarvestani}, {Scarlata},
  {Schinnerer}, {Sheth}, {Shopbell}, {Smol{\v c}i{\'c}}, {Taniguchi}, {Taylor},
  {White}, \& {Yan}}]{Scoville13}
{Scoville}, N., {Arnouts}, S., {Aussel}, H., {et~al.} 2013, \apjs, 206, 3

\bibitem[{{Shull} {et~al.}(2012){Shull}, {Smith}, \& {Danforth}}]{Shull12}
{Shull}, J.~M., {Smith}, B.~D., \& {Danforth}, C.~W. 2012, \apj, 759, 23

\bibitem[{{Smith} {et~al.}(2012){Smith}, {Hopkins}, {Hunstead}, \&
  {Pimbblet}}]{Smith12}
{Smith}, A.~G., {Hopkins}, A.~M., {Hunstead}, R.~W., \& {Pimbblet}, K.~A. 2012,
  \mnras, 422, 25

\bibitem[{{Snedden} {et~al.}(2016){Snedden}, {Coughlin}, {Phillips}, {Mathews},
  \& {Suh}}]{Snedden16}
{Snedden}, A., {Coughlin}, J., {Phillips}, L.~A., {Mathews}, G., \& {Suh},
  I.-S. 2016, \mnras, 455, 2804

\bibitem[{{Snedden} {et~al.}(2015){Snedden}, {Phillips}, {Mathews}, {Coughlin},
  {Suh}, \& {Bhattacharya}}]{Snedden15}
{Snedden}, A., {Phillips}, L.~A., {Mathews}, G.~J., {et~al.} 2015, Journal of
  Computational Physics, 299, 92

\bibitem[{{Sobral} {et~al.}(2011){Sobral}, {Best}, {Smail}, {Geach},
  {Cirasuolo}, {Garn}, \& {Dalton}}]{Sobral11}
{Sobral}, D., {Best}, P.~N., {Smail}, I., {et~al.} 2011, \mnras, 411, 675

\bibitem[{{Sobral} {et~al.}(2015){Sobral}, {Stroe}, {Dawson}, {Wittman}, {Jee},
  {R{\"o}ttgering}, {van Weeren}, \& {Br{\"u}ggen}}]{Sobral15}
{Sobral}, D., {Stroe}, A., {Dawson}, W.~A., {et~al.} 2015, \mnras, 450, 630

\bibitem[{{Sobral} {et~al.}(2016){Sobral}, {Stroe}, {Koyama}, {Darvish},
  {Calhau}, {Afonso}, {Kodama}, \& {Nakata}}]{Sobral16}
{Sobral}, D., {Stroe}, A., {Koyama}, Y., {et~al.} 2016, \mnras, 458, 3443

\bibitem[{{Sousbie}(2011)}]{Sousbie11}
{Sousbie}, T. 2011, \mnras, 414, 350

\bibitem[{{Sousbie} {et~al.}(2008){Sousbie}, {Pichon}, {Colombi}, {Novikov}, \&
  {Pogosyan}}]{Sousbie08}
{Sousbie}, T., {Pichon}, C., {Colombi}, S., {Novikov}, D., \& {Pogosyan}, D.
  2008, \mnras, 383, 1655

\bibitem[{{Stoica} {et~al.}(2005){Stoica}, {Mart{\'{\i}}nez}, {Mateu}, \&
  {Saar}}]{Stoica05}
{Stoica}, R.~S., {Mart{\'{\i}}nez}, V.~J., {Mateu}, J., \& {Saar}, E. 2005,
  \aap, 434, 423

\bibitem[{{Stoica} {et~al.}(2010){Stoica}, {Mart{\'{\i}}nez}, \&
  {Saar}}]{Stoica10}
{Stoica}, R.~S., {Mart{\'{\i}}nez}, V.~J., \& {Saar}, E. 2010, \aap, 510, A38

\bibitem[{{Stroe} {et~al.}(2015){Stroe}, {Sobral}, {Dawson}, {Jee}, {Hoekstra},
  {Wittman}, {van Weeren}, {Br{\"u}ggen}, \& {R{\"o}ttgering}}]{Stroe15}
{Stroe}, A., {Sobral}, D., {Dawson}, W., {et~al.} 2015, \mnras, 450, 646

\bibitem[{{Tempel} \& {Libeskind}(2013)}]{Tempel13b}
{Tempel}, E., \& {Libeskind}, N.~I. 2013, \apjl, 775, L42

\bibitem[{{Tempel} {et~al.}(2014){Tempel}, {Stoica}, {Mart{\'{\i}}nez},
  {Liivam{\"a}gi}, {Castellan}, \& {Saar}}]{Tempel14}
{Tempel}, E., {Stoica}, R.~S., {Mart{\'{\i}}nez}, V.~J., {et~al.} 2014, \mnras,
  438, 3465

\bibitem[{{Tempel} {et~al.}(2013){Tempel}, {Stoica}, \& {Saar}}]{Tempel13a}
{Tempel}, E., {Stoica}, R.~S., \& {Saar}, E. 2013, \mnras, 428, 1827

\bibitem[{{Tran} {et~al.}(2015){Tran}, {Nanayakkara}, {Yuan}, {Kacprzak},
  {Glazebrook}, {Kewley}, {Momcheva}, {Papovich}, {Quadri}, {Rudnick},
  {Saintonge}, {Spitler}, {Straatman}, \& {Tomczak}}]{Tran15}
{Tran}, K.-V.~H., {Nanayakkara}, T., {Yuan}, T., {et~al.} 2015, \apj, 811, 28

\bibitem[{{Trowland} {et~al.}(2013){Trowland}, {Lewis}, \&
  {Bland-Hawthorn}}]{Trowland13}
{Trowland}, H.~E., {Lewis}, G.~F., \& {Bland-Hawthorn}, J. 2013, \apj, 762, 72

\bibitem[{{Vogelsberger} {et~al.}(2014){Vogelsberger}, {Genel}, {Springel},
  {Torrey}, {Sijacki}, {Xu}, {Snyder}, {Bird}, {Nelson}, \&
  {Hernquist}}]{Vogelsberger14}
{Vogelsberger}, M., {Genel}, S., {Springel}, V., {et~al.} 2014, \nat, 509, 177

\bibitem[{{von der Linden} {et~al.}(2010){von der Linden}, {Wild}, {Kauffmann},
  {White}, \& {Weinmann}}]{Vonderlinden10}
{von der Linden}, A., {Wild}, V., {Kauffmann}, G., {White}, S.~D.~M., \&
  {Weinmann}, S. 2010, \mnras, 404, 1231

\bibitem[{{Vulcani} {et~al.}(2010){Vulcani}, {Poggianti}, {Finn}, {Rudnick},
  {Desai}, \& {Bamford}}]{Vulcani10}
{Vulcani}, B., {Poggianti}, B.~M., {Finn}, R.~A., {et~al.} 2010, \apjl, 710, L1

\bibitem[{{Vulcani} {et~al.}(2016{\natexlab{a}}){Vulcani}, {Treu}, {Schmidt},
  {Morishita}, {Dressler}, {Poggianti}, {Abramson}, {Brada{\v c}}, {Brammer},
  {Hoag}, {Malkan}, {Pentericci}, \& {Trenti}}]{Vulcani16b}
{Vulcani}, B., {Treu}, T., {Schmidt}, K.~B., {et~al.} 2016{\natexlab{a}}, ArXiv
  e-prints, arXiv:1610.04621

\bibitem[{{Vulcani} {et~al.}(2016{\natexlab{b}}){Vulcani}, {Treu}, {Nipoti},
  {Schmidt}, {Dressler}, {Morshita}, {Poggianti}, {Malkan}, {Hoag}, {Brada{\v
  c}}, {Abramson}, {Trenti}, {Pentericci}, {von der Linden}, {Morris}, \&
  {Wang}}]{Vulcani16}
{Vulcani}, B., {Treu}, T., {Nipoti}, C., {et~al.} 2016{\natexlab{b}}, ArXiv
  e-prints, arXiv:1610.04615

\bibitem[{{Wang} {et~al.}(2012){Wang}, {Mo}, {Yang}, \& {van den
  Bosch}}]{Wang12}
{Wang}, H., {Mo}, H.~J., {Yang}, X., \& {van den Bosch}, F.~C. 2012, \mnras,
  420, 1809

\bibitem[{{Wang} {et~al.}(2016){Wang}, {Mo}, {Yang}, {Zhang}, {Shi}, {Jing},
  {Liu}, {Li}, {Kang}, \& {Gao}}]{Wang16}
{Wang}, H., {Mo}, H.~J., {Yang}, X., {et~al.} 2016, ArXiv e-prints,
  arXiv:1608.01763

\bibitem[{{Weinmann} {et~al.}(2006){Weinmann}, {van den Bosch}, {Yang}, \&
  {Mo}}]{Weinmann06}
{Weinmann}, S.~M., {van den Bosch}, F.~C., {Yang}, X., \& {Mo}, H.~J. 2006,
  \mnras, 366, 2

\bibitem[{{Welker} {et~al.}(2015){Welker}, {Dubois}, {Pichon}, {Devriendt}, \&
  {Chisari}}]{Welker15}
{Welker}, C., {Dubois}, Y., {Pichon}, C., {Devriendt}, J., \& {Chisari}, E.~N.
  2015, ArXiv e-prints, arXiv:1512.00400

\bibitem[{{Wetzel} {et~al.}(2013){Wetzel}, {Tinker}, {Conroy}, \& {van den
  Bosch}}]{Wetzel13}
{Wetzel}, A.~R., {Tinker}, J.~L., {Conroy}, C., \& {van den Bosch}, F.~C. 2013,
  \mnras, 432, 336

\bibitem[{{Wetzel} {et~al.}(2014){Wetzel}, {Tinker}, {Conroy}, \& {van den
  Bosch}}]{Wetzel14}
---. 2014, \mnras, 439, 2687

\bibitem[{{White}(1984)}]{White84}
{White}, S.~D.~M. 1984, \apj, 286, 38

\bibitem[{{Wijesinghe} {et~al.}(2012){Wijesinghe}, {Hopkins}, {Brough},
  {Taylor}, {Norberg}, {Bauer}, {Brown}, {Cameron}, {Conselice}, {Croom},
  {Driver}, {Grootes}, {Jones}, {Kelvin}, {Loveday}, {Pimbblet}, {Popescu},
  {Prescott}, {Sharp}, {Baldry}, {Sadler}, {Liske}, {Robotham}, {Bamford},
  {Bland-Hawthorn}, {Gunawardhana}, {Meyer}, {Parkinson}, {Drinkwater},
  {Peacock}, \& {Tuffs}}]{Wijesinghe12}
{Wijesinghe}, D.~B., {Hopkins}, A.~M., {Brough}, S., {et~al.} 2012, \mnras,
  423, 3679

\bibitem[{{Zel'dovich}(1970)}]{Zeldovich70}
{Zel'dovich}, Y.~B. 1970, \aap, 5, 84

\bibitem[{{Zhang} {et~al.}(2009){Zhang}, {Yang}, {Faltenbacher}, {Springel},
  {Lin}, \& {Wang}}]{Zhang09}
{Zhang}, Y., {Yang}, X., {Faltenbacher}, A., {et~al.} 2009, \apj, 706, 747

\bibitem[{{Zhang} {et~al.}(2013){Zhang}, {Yang}, {Wang}, {Wang}, {Mo}, \& {van
  den Bosch}}]{Zhang13}
{Zhang}, Y., {Yang}, X., {Wang}, H., {et~al.} 2013, \apj, 779, 160

\end{thebibliography}

\end{document}